\documentclass{LMCS}
\usepackage{enumerate}
\usepackage{amssymb}
\usepackage{amsmath}
\usepackage{latexsym}
\usepackage{bussproofs}
\usepackage{hyperref}

\newenvironment{DEF}{\begin{defi}}{\end{defi}}
\newenvironment{REM}{\begin{rem}}{\end{rem}} 
\newenvironment{expl}{\begin{exa}}{\end{exa}}
\newenvironment{Th}{\begin{thm}}{\end{thm}}

\def\varx{x}

\def\subst#1#2{\left\lbrack{#2}\mathrm{:=}{#1}\right\rbrack}

\def\land{\wedge} 
\def\lor{\vee}    

\def\NN{\mathbb{N}}

\def\BNFdef{\mathtt{~::=~}}
\def\BNFor{\mid}

\def\ll{\lambda\kern-4.3pt\lambda}       
\def\xx{x\kern-4pt x}                    

\def\red{\to}           

\def\vec#1{\overrightarrow{#1\,\,}}  

\def\EX{\mathsf{EX}}
\def\AX{\mathsf{AX}}
\def\EG{\mathsf{EG}}
\def\AG{\mathsf{AG}}
\def\ll{\mathcal L}
\def\restrict#1{{\mid}_{#1}}

\def\A{\mathfrak A}

\def\pow#1{\mathfrak P({#1})}

\newcommand{\monus}{\mathbin{-\kern-.51em{^.}\kern.51em}}
\def\Cdelay{\mathcal W}
\def\rep{\mathcal R}
\def\app{\mathtt{@}}
\def\red{\to_\beta}
\def\lang#1{\mathcal L({#1})}

\def\runI#1{\mbox{$\A\models^{\infty}{#1}$}}
\def\runIn#1#2{\mbox{$\A\models^{#1}{#2}$}}
\def\run#1#2#3{\mbox{$\A,{#2}\models^{#1}{#3}$}}

\def\CN#1#2{{#1}{\underline{\mathbf @}}{#2}}
\def\term{\mathfrak f}
\def\termg{\mathfrak g}
\def\ar#1{\sharp({#1})}
\def\sem#1{[\!\![{#1}]\!\!]}
\def\less{\sqsubseteq}
\def\fproof#1#2#3#4#5{\mbox{${#2}\vdash^{#1}_\A {#3}\sqsubseteq{#4}:{#5}$}}

\def\force#1#2#3#4{\mbox{${#2}\prec\!\!\!\prec^{#1}_\A{#3}:{#4}$}}

\def\subst#1#2{[{#1}/{#2}]}
\def\dom#1{\mathrm{dom}({#1})}

\def\csem#1{\langle\!\langle{#1}\rangle\!\rangle_{\A\infty}}
\def\csemC#1#2{\langle\!\langle{#2}\rangle\!\rangle_{\A\infty}^{#1}}

\def\qeven{q_2}
\def\qodd{q_1}
\def\id{\mathrm{id}}
\def\swap{\mathrm{swap}}
\def\Tsim#1{\approx_{#1}}

\def\doi{3 (3:1) 2007}
\lmcsheading%
{\doi}
{1--23}
{}
{}
{Nov.~24, 2006}
{Jul.~\phantom{0}4, 2007}
{}   

\begin{document}
\title[A Finite Semantics\dots\ for Infinite Runs of Automata]{A
  Finite Semantics of Simply-Typed Lambda Terms\\ for Infinite Runs of Automata}

\author[K.~Aehlig]{Klaus Aehlig}
\thanks{Partially supported by grant EP/D03809X/1 of the British 
Engineering and Physical Sciences Research Council (EPSRC). Part of
this article was written while Klaus Aehlig was affiliated with the
University of Toronto and supported by grant Ae 102-1/1 of the
``Deutsche Forschungsgemeinschaft'' (DFG)}
\address{Department of Computer Science\\University of Wales
  Swansea\\Swansea SA2 8PP\\United Kingdom}
\email{k.t.aehlig@swan.ac.uk}
\titlecomment{}
\begin{abstract}
Model checking properties are often described by means of finite
automata. Any particular such automaton divides the set of infinite trees into
finitely many classes, according to which state has an infinite
run. Building the full type hierarchy upon this interpretation of the
base type gives a finite semantics for simply-typed lambda-trees.

A calculus based on this semantics is proven sound and complete. In
particular, for regular infinite lambda-trees it is decidable whether
a given automaton has a run or not. As regular lambda-trees are
precisely recursion schemes, this decidability result holds for
arbitrary recursion schemes of arbitrary level, without any
syntactical restriction.
\end{abstract}
\maketitle

{
\def\Myvec#1{\vec{#1}}

\begin{section}{Introduction and Related Work}

The lambda calculus~\cite{Barendregt77} has long been used as model of
computation. In its untyped form it is Turing complete. Even though
models of the untyped lambda calculus are known, restricting it to a
typing discipline allows for more specific models. The simply-typed lambda
calculus has a straight forward set-theoretic semantics.

Quite early on, not only finite but also infinite lambda-terms have
been considered. For example, Barendregt~\cite{Barendregt77}
introduced the concept of ``B\"ohm trees'' as a generalised concept of
normal forms for lambda-terms where normalisation does not necessarily
terminate, but still might produce a growing normal prefix; for
example the term $Y(\lambda zx.xz)$ has the B\"ohm tree $\lambda
x. x(\lambda x. x(\lambda x. x\ldots))$.

Since Rabin~\cite{Rabin69} showed the decidability of the monadic second
order (MSO) theory of the infinite binary tree this result has been applied and
extended to various mathematical structures, including algebraic
trees~\cite{Courcelle95} and a hierarchy of graphs~\cite{Caucal96}
obtained by iterated unfolding and inverse rational mappings from
finite graphs. The interest in these kind of structures arose in recent
years in the context of verification of infinite state
systems~\cite{KupfermanVardi00,Walukiewicz01}.

Recently Knapik, Niwi\'nski and
Urzyczyn~\cite{KnapikNiwinskiUrzyczyn01} showed that the monadic
second order
theory of any infinite tree generated by a level-$2$ grammar
satisfying a certain ``safety'' condition is decidable.
Later they
generalised~\cite{KnapikNiwinskiUrzyczyn02} this result to grammars of
arbitrary levels, but still requiring the ``safety'' condition. In
particular, the question was left open whether a ``safety'' constraint is necessary
to obtain decidability. In this article we will give a partial answer.

It should be noted that trees given by higher-order grammars can also
be understood as trees given by simply-typed infinite, but regular,
lambda terms. The ``safety'' condition guarantees that beta-reduction
can be carried out in such a way that variables never have to be
renamed in the process of substitution. This obviously is a property
related to operational aspects of computation. Our approach to avoid
the need for such a restriction is therefore to search for a
\emph{denotational} semantics. Denotational approaches tend to be less
vulnerable to the need of requiring specific \emph{operational} properties.

To obtain effective constructions, like an effective semantics, it is
useful to have a concrete representation of the properties to be
verified. Finite automata are a standard tool to do so. In this
article we concentrate on automata with trivial acceptance
condition. These automata do not exhaust the full of MSO but, as we
shall see, are able to express a reasonable set of safety
properties.

Their advantage, however, is that they seem particularly suited for a
denotational approach. The reason is, that the ``interface'' is
particularly simple. In order to combine two partial runs into a longer
run, the only thing we have to look at is the state in which the
automaton arrives. 

Based on this intuition we construct a semantics for the simple
types. Actually, we use the standard set-theoretic semantics. Hence
the only thing we have to specify is the interpretation of the base
type. Following the discussion above, we describe a term of base type
by the set of states a given automaton can start a run on the tree
denoted by that term. 

More precisely, we consider the following problem.
\medskip\begin{quote}
Given a, possibly infinite, simply-typed lambda-tree $t$ of base type,
and given a non-deterministic tree automaton $\A$. Does $\A$ have a
run on the normal form of $t$?
\end{quote}\medskip

\noindent The idea is to provide a ``proof'' of a run of $\A$ on the
normal form of $t$ by annotating each subterm of $t$ with a semantical
value describing how this subterm ``looks, as seen by $\A$''. Since,
in the end, all the annotations come from a fixed finite set, the
existence of such a proof is decidable.

The idea of a ``proof'' that a given automaton has a run on a
tree is used, at least implicitly, in the work by Aehlig, de Miranda
and Ong~\cite{AehligMirandaOng04}. This work also gives an affirmative
answer to the question of the decidability for the full MSO theory for
trees generated by level-two recursion schemes.

Very recently, simultaneously and independently, Luke Ong could give
an affirmative answer~\cite{Ong06} for trees generated by recursion schemes of
arbitrary level, still deciding the full MSO theory; he thus obtained
a stronger result in what concerns decidability.
His result is based on game semantics~\cite{HylandOng00} and
is technically quite involved. Therefore the author believes that his
conceptually more simple approach still is of worth. Moreover, the
novel finitary semantics for the simple types introduced in
this article, and the sound and complete proof system to show the
existence of a run of an automaton seem to be of independent interest.
An extended abstract~\cite{Aehlig06} of this article appeared in the
proceedings of CSL~'06.

This article is organised as follows. In Section~\ref{sEc:autom-new}
we formally introduce automata with trivial acceptance condition and
study their languages. We also prove the closure of these languages
under the modality ``globally''. We also show that properties based on 
the modality ``eventually'' are not
expressible. In Section~\ref{sEc:inf-lam-trees} we introduce
infinitary simply-typed lambda trees and in
Section~\ref{sEc:rec-scheme-def} we introduce recursion schemes
as a means to describe regular lambda
trees. This also shows that some lambda trees have a representation
that is not only effective, but also quite natural. In
Section~\ref{sEc:cn} we explain continuous normalisation for the lambda
calculus. The use of continuous normalisation is twofold. On the one
hand, it allows simpler definitions and proofs, as one layer of input
corresponds precisely to one layer of output. On the other hand, it
is simply a necessity in order to have a well-defined normal form in the
presence of non-terminating computations due to the infinitary nature
of our lambda trees. Section~\ref{sEc:fin-sem} introduces the finitary
semantics and the proof system; Sections~\ref{sEc:sound}
and~\ref{sEc:complete} are devoted to
the proofs of its soundness and completeness.
Finally, in Section~\ref{sEc:model-checking}, we put the results
together to obtain the mentioned decidability result.

\end{section}
\begin{section}{Automata with Trivial Acceptance
	Condition}\label{sEc:autom-new}
We assume a set of {\em letters} or {\em terminals} be given to us as
a primitive notion. We use $\term$ to range over letters.
Each letter $\term$ is associated an {\it arity} $\ar\term\in\NN$. 

\begin{DEF}
For $\Sigma$ a set of terminals, a \emph{$\Sigma$-term} is a, not necessarily
well-founded, tree labelled with elements of $\Sigma$ where every node
labelled with $\term$ has $\ar\term$ many children.

A \emph{$\Sigma$-language} is any subset of the set of all
$\Sigma$-terms. We use the term \emph{language} if $\Sigma$ is
understood.
\end{DEF}

\begin{expl}\label{expl:tree}
Let $\Sigma'=\{\mathtt f,\mathtt g,\mathtt a\}$ with $\mathtt f$,
$\mathtt g$ and $\mathtt a$ of arities $2$, $1$, and
$0$, respectively. 
Figure~\ref{fig:valuetree} shows two $\Sigma'$-terms.
\end{expl}

\begin{figure}
{
\setlength\unitlength{12pt}
\begin{picture}(6,11)(-2,-10)

\put (-0.5,0) {\makebox[1\unitlength]{$\mathtt f$}}
\put (0,0) {\line(-1,-1) 1}
\put (0,0) {\line(1,-1) 1}
\put (-1.5,-1.7) {\makebox[1\unitlength]{$\mathtt a$}}
\put (0.5,-2) {\makebox[1\unitlength]{$\mathtt f$}}
\put (1,-2) {\line (-1,-1) 1}
\put (1,-2) {\line (1,-1) 1}
\put (-0.5,-3.7) {\makebox[1\unitlength]{$\mathtt g$}}
\put (1.5,-4) {\makebox[1\unitlength]{$\mathtt f$}}
\put (0,-4) {\line(0,-1) 1}
\put (-0.5,-5.7) {\makebox[1\unitlength]{$\mathtt a$}}
\put (2,-4) {\line (-1,-1) 1}
\put (2,-4) {\line (1,-1) 1}

\put (0.5,-5.7) {\makebox[1\unitlength]{$\mathtt g$}}
\put (1,-6) {\line(0,-1) 1}
\put (0.5,-7.7) {\makebox[1\unitlength]{$\mathtt g$}}
\put (1,-8) {\line(0,-1) 1}
\put (0.5,-9.7) {\makebox[1\unitlength]{$\mathtt a$}}

\put (2.5,-6) {\makebox[1\unitlength]{$\vdots$}}
\end{picture}%
}%
\quad%
{
\setlength\unitlength{12pt}
\begin{picture}(6,19)(-2,-18)

\put (-0.5,0) {\makebox[1\unitlength]{$\mathtt f$}}
\put (0,0) {\line(-1,-1) 1}
\put (0,0) {\line(1,-1) 1}
\put (-1.5,-1.7) {\makebox[1\unitlength]{$\mathtt g$}}
\put (-1,-2) {\line(0,-1) 1}
\put (-1.5,-3.7) {\makebox[1\unitlength]{$\mathtt g$}}
\put (-1,-4) {\line(0,-1) 1}
\put (-1.5,-5.7) {\makebox[1\unitlength]{$\mathtt a$}}

\put (0.5,-2) {\makebox[1\unitlength]{$\mathtt f$}}
\put (1,-2) {\line (-1,-1) 1}
\put (1,-2) {\line (1,-1) 1}
\put (-0.5,-3.7) {\makebox[1\unitlength]{$\mathtt g$}}
\put (0,-4) {\line(0,-1) 1}
\put (-0.5,-5.7) {\makebox[1\unitlength]{$\mathtt g$}}
\put (0,-6) {\line(0,-1) 1}
\put (-0.5,-7.7) {\makebox[1\unitlength]{$\mathtt g$}}
\put (0,-8) {\line(0,-1) 1}
\put (-0.5,-9.7) {\makebox[1\unitlength]{$\mathtt g$}}
\put (0,-10) {\line(0,-1) 1}
\put (-0.5,-11.7) {\makebox[1\unitlength]{$\mathtt a$}}

\put (1.5,-4) {\makebox[1\unitlength]{$\mathtt f$}}
\put (2,-4) {\line (-1,-1) 1}
\put (2,-4) {\line (1,-1) 1}

\put (0.5,-5.7) {\makebox[1\unitlength]{$\mathtt g$}}
\put (1,-6) {\line(0,-1) {0.5}}
\put (0.5,-7.2) {\makebox[1\unitlength]{$\mathtt g$}}
\put (1,-7.5) {\line(0,-1) {0.5}}
\put (0.5,-8.7) {\makebox[1\unitlength]{$\mathtt g$}}
\put (1,-9) {\line(0,-1) {0.5}}
\put (0.5,-10.2) {\makebox[1\unitlength]{$\mathtt g$}}
\put (1,-10.5) {\line(0,-1) {0.5}}
\put (0.5,-11.7) {\makebox[1\unitlength]{$\mathtt g$}}
\put (1,-12) {\line(0,-1) {0.5}}
\put (0.5,-13.2) {\makebox[1\unitlength]{$\mathtt g$}}
\put (1,-13.5) {\line(0,-1) {0.5}}
\put (0.5,-14.7) {\makebox[1\unitlength]{$\mathtt g$}}
\put (1,-15) {\line(0,-1) {0.5}}
\put (0.5,-16.2) {\makebox[1\unitlength]{$\mathtt g$}}
\put (1,-16.5) {\line(0,-1) {0.5}}
\put (0.5,-17.7) {\makebox[1\unitlength]{$\mathtt a$}}

\put (2.5,-6) {\makebox[1\unitlength]{$\vdots$}}
\end{picture}%
}%

\caption{Two $\{\mathtt f,\mathtt g,\mathtt a\}$-terms.\label{fig:valuetree}}
\end{figure}

\begin{DEF}[Trivial Automata]
A non-deterministic tree automaton with trivial acceptance condition over
the alphabet $\Sigma$, or a ``trivial automaton'' for short, is given by
\begin{enumerate}[$\bullet$]
\item a finite set $Q$ of ``states'',
\item a set $I\subset Q$ of ``initial states'', and
\item a transition function $\delta\colon Q\times\Sigma\to\pow{(Q\cup\{\ast\})^N}$.
\end{enumerate}
Here $N=\max\{\ar\termg\mid\termg\in\Sigma\}$ is the maximal arity and
we require
$\delta(q,\termg)\subset Q^{\ar\termg}\times\{\ast\}^{N-\ar\termg}$
whenever $q\in Q$ and $\termg\in\Sigma$. 
\end{DEF}

\begin{DEF}[Run of a Trivial Automaton]
If $t$ is $\Sigma$-term, and $\A$ a trivial automaton over $\Sigma$,
then a \emph{run} (also ``an infinite run'') \emph{of $\A$ on $t$ starting
in state $q$} is a mapping $r$ from the nodes of $t$ to $Q$, such that the root is mapped
to $q$, and, whenever $p$ is a $\term$-labelled node in
$t$ and $p_1,\ldots,p_{\ar\term}$ are the children of $p$, then
$(r(p_1),\ldots,r(p_{\ar\term}),\ast,\ldots,\ast)\in\delta(r(p),\term)$. 

A \emph{run up to level $n$ starting in state $q$} is a mapping from
all nodes of $t$ with distance at most $n$ to $Q$ such that the
above condition holds for all nodes $p,\vec p$ in the domain of $r$,
i.e., whenever a node $p$ is $\term$-labelled and its children
$p_1,\ldots,p_{\ar\term}$ have distance at most $n$ to the root,
then $(r(p_1),\ldots,r(p_{\ar\term}),\ast,\ldots,\ast)\in\delta(r(p),\term)$. 

A \emph{run} or a \emph{run up to level $n$}, is a run or a run up to
level $n$ starting in some initial state.
\end{DEF}

We write $\run nqt$ to denote that $\A$ has a run on $t$ up to level $n$
starting in state $q$. We write $\run \infty q t$ to denote that $\A$
has a run on $t$ starting in state $q$. We write $\runIn n t$ to
denote that $\A$ has a run up to level $n$ on $t$ and 
we write $\runI t$ to denote that $\A$ has a run on $t$.

\begin{REM}
Trivially, every automaton has a run up to level $0$ on every
term starting in every state. Also immediate from the definition we
see that, if $\A$ has a run up to level $n$ on $t$ and $m\leq n$ then
$\A$ has a run up to level $m$ on $t$. 
\end{REM}

\begin{REM}
By K\"onig's Lemma $\A$ has a run on $t$ if and only if $\A$ has a run
up to level $n$ on $t$ for every $n\in\NN$.
\end{REM}

\begin{expl}\label{expl:prop}
Continuing Example~\ref{expl:tree}
consider the property
\begin{quote}
``Every maximal chain of letters $\mathtt g$ has even length''.
\end{quote}
It can be
expressed by an automaton with two states $Q=\{\qeven,\qodd\}$ where
$\qeven$  means that an even number of $\mathtt g$s has been passed on the
path so far, and $\qodd$ means that the maximal chain of $\mathtt g$s passed
has odd length. Then the initial state is $\qeven$ and the transition
function is as follows.
$$
\begin{array}{lcllcl}
\delta(\mathtt f,\qeven)&=&\{(\qeven,\qeven)\}\qquad &\delta(\mathtt f,\qodd) &=& \emptyset \\
\delta(\mathtt g,\qeven)&=&\{(\qodd,\ast)\}\qquad & \delta(\mathtt g,\qodd)&=&\{(\qeven,\ast)\}\\
\delta(\mathtt a,\qeven)&=&\{(\ast,\ast)\}\qquad  & \delta(\mathtt a,\qodd) &=& \emptyset
\end{array}
$$
Note that this automaton has an infinite run on the second tree in
Figure~\ref{fig:valuetree}, whereas it has a run only up to level $3$
on the first one.
\end{expl}

\begin{DEF}[$\lang\A$]
If $\A$ is a trivial automaton over the alphabet $\Sigma$ then by
$\lang\A$ we denote the language of $\A$, that is, the set
$$\lang\A=\{t\mid\runI t\}$$
of all terms $t$ such that $\A$ has a run on $t$.
\end{DEF}

\begin{prop}\label{prop:trivial-autom:letter}
There exists a trivial automaton that accepts a tree if and only if
its root is labelled by the terminal $\term$.
\end{prop}
\begin{proof}
Let $q_1$ be an all-accepting state, i.e.,
$\delta(q_1,\termg)=\{(q_1,\ldots,q_1,\ast,\ldots,\ast)\}$ for all
$\termg\in\Sigma$.
Let $q_0$ be the only initial state, and set
$\delta(q_0,\term)=\{(q_1,\ldots,q_1,\ast,\ldots,\ast)\}$ and
$\delta(q_0,\termg)=\emptyset$ for $\termg\neq\term$.
\end{proof}

\begin{lem}\label{lem:trivial-autom:disj}
If $\A_0$ and $\A_1$ are trivial automata, then there is a trivial
automaton $\A$ with
$\lang\A=\lang{\A_0}\cup\lang{\A_1}$.
\end{lem}
\begin{proof}
Let $\A_i$ have state set $Q_i$, initial states $I_i$ and transition
$\delta_i$. Assume, without loss of generality, that $Q_0$ and $Q_1$
are disjoint. Then $\A$ is given by the following data. State set is
$Q=Q_0\cup Q_1$, initial states are $I=I_0\cup I_1$ and the transition
function $\delta$ is defined by $\delta(q,\term)=\delta_i(q,\term)$ for $q\in Q_i$.
\end{proof}

\begin{lem}\label{lem:trivial-autom:conj}
If $\A_0$ and $\A_1$ are trivial automata, then there is a trivial
automaton $\A$ with
$\lang\A=\lang{\A_0}\cap\lang{\A_1}$.
\end{lem}
{\sloppy
\begin{proof}
Let $\A_i$ have state set $Q_i$, initial states $I_i$ and transition
$\delta_i$. Set $Q=Q_0\times Q_1$, $I=I_0\times I_1$ and define
$\delta \colon(Q_0\times Q_1)\times 
\Sigma\to\pow{(Q_0\times Q_1\cup\{\ast\})^N}$ by
$\delta((q,q'),\term) =
\{((q_0,q_0'),\ldots,(q_{\ar\term},q_{\ar\term}'),\ast,\ldots,\ast) 
\mid (q_0,\ldots,q_{\ar\term},\ldots)\in\delta_0(q,\term)
\land(q_0',\ldots,q_{\ar\term}',\ldots)\in\delta_1(q,\term) \}$.
Then $Q$, $I$ and $\delta$ define an automaton $\A$ as desired.
\end{proof}
}


Non-determinism immediately provides us with closure under projection
of the alphabet; we'll give a precise definition of this property.

\begin{DEF}\label{def:proj-alphabet-terms}
If $\Sigma$ and $\overline\Sigma$ are sets of terminals, a
\emph{projection from $\Sigma$ to $\overline\Sigma$}, is a mapping
$\pi\colon\Sigma\to\overline\Sigma$ such that
$\ar{\pi(\term)}=\ar\term$ for all $\term\in\Sigma$. If $t$ is a
$\Sigma$-term and $\pi$ is a projection from $\Sigma$ to
$\overline\Sigma$, then by $\pi(t)$ we denote the
$\overline\Sigma$-term that is obtained from $t$ by replacing every
label $\term$ by $\pi(\term)$.
\end{DEF}

\begin{REM}
In Definition~\ref{def:proj-alphabet-terms} the condition on the arity
is necessary to ensure that $\pi(t)$ is a well-formed
$\overline\Sigma$-tree, i.e., every node $\termg$-labelled node has
$\ar\termg$ many children.
\end{REM}

\begin{lem}
If $\Sigma$ and $\overline\Sigma$ are sets of terminals,
$\pi$ is a projection from $\Sigma$ to $\overline\Sigma$, and
$\A$ is a trivial automaton $\Sigma$, then there is a trivial
automaton $\A_\pi$ such that
$$\lang{\A_\pi}=\{ \pi(t)\mid t\in\lang \A \}\;.$$
\end{lem}
\begin{proof}
Let $\A$ have state set $Q$, initial states $I$ and transition
$\delta$. Then a possible automaton $\A_\pi$ is given by the same set $Q$
of states
and the same set $I$ of initial state, but with transition function
$\delta_\pi$ defined by $\delta_\pi(q,\termg) =
\bigcup\{\delta(q,\term)\mid \term\in\Sigma,\pi(\term)=\termg\}$. 
\end{proof}

Another obvious closure property of the languages of trivial automata
are the temporal ``next'' operators.

\begin{DEF}[$\EX\ll$, $\AX\ll$]
If $\ll$ is a language we define the languages
$$\EX\ll=\{\term t_1\ldots t_{\ar\term}\mid \exists i. t_i\in\ll\}$$
and
$$\AX\ll=\{\term t_1\ldots t_{\ar\term}\mid \forall i. t_i\in\ll\}\;.$$
\end{DEF}

\begin{lem}\label{lem:trivial-autom:next}
If $\A$ is a trivial automaton, then there exist trivial automata
$\A_\EX$ and $\A_\AX$ with $\lang{\A_\EX}=\EX\lang\A$ and $\lang{\A_\AX}=\AX\lang\A$.
\end{lem}
\begin{proof}
To construct $\A_\AX$, add a new state $q_0$ to the state set of
$\A$. This new state will be the only initial state of
$\A_\AX$. Extend the transition function $\delta$ by setting
$$\delta(q_0,\term)=\{(q_1,\ldots,q_{\ar\term},\ast,\ldots,\ast) |
q_1,\ldots,q_{\ar\term}\in I\}$$ 
where $I$ is the set of initial states of $\A$.

To construct $\A_\EX$ from $\A$ add a new state $q_0$, which will be the only
initial state of the new automaton, and add a new all-accepting state
$q_f$. Extend $\delta$ by setting
$$\delta(q_0,\term)=\{
\begin{array}[t]{l}(q_i,q_f\ldots,q_f,q_f,\ast,\ldots,\ast),\\
(q_f,q_i\ldots,q_f,q_f,\ast,\ldots,\ast),\\\ldots\\
(q_f,q_f\ldots,q_i,q_f,\ast,\ldots,\ast),\\
(q_f,q_f\ldots,q_f,q_i,\ast,\ldots,\ast)\; | q_i\in I\}
\end{array}
$$ 
where $I$ is the set of initial states of $\A$.
\end{proof}

\begin{DEF}[$p\in t$, $t\restrict p$, Path]
We use $p\in t$ to express that $p$ is a node in $t$. In this case we
write $t\restrict p$ for the subterm of $t$ whose root is $p$.

A \emph{path} in $t$ is a maximal set $P$ of nodes in $t$ such that if a node
$p\in t$ different from the root is in $P$, then so is its parent, and
such that for every node in $P$ at most one of its children is in $P$.
\end{DEF}

\begin{REM}
Immediately from the definition of a path we note that if $P$ is a
path in $t$ and  $p\in P$ has a child in $t$ then some child of
$p$ has to be in $P$.
\end{REM}

\begin{DEF}
If $\ll$ is a language we define the languages
$$\EG\ll=\{t\mid\exists P(\text{$P$ path in $t$}\land\forall p\in
P. t\restrict p\in\ll)\}$$
and
$$\AG\ll=\{t\mid\forall p\in t. t\restrict p\in\ll\}\;.$$
\end{DEF}

The next lemma states that the set of languages of trivial automata is
closed under the modal operator ``globally''. On the one hand, this is an
interesting closure property, which shows that at least safety
properties can be expressed by trivial automata. On the other hand, it
is worth looking at the proof of this lemma, as it shows, in a simple
setting, all the central ideas that will be used to construct our
finitary proof calculus and show its soundness and completeness. The
states of the automaton $\A_\AG$ constructed in the proof of
Lemma~\ref{lem:trivial-autom:gloabally} should be thought of as
annotations proving that $\A$ has a run starting in various states.

\begin{lem}\label{lem:trivial-autom:gloabally}
If $\A$ is a trivial automaton, then there exist trivial automata
$\A_\EG$ and $\A_\AG$ such that $\lang{\A_\EG}=\EG\lang\A$ and
$\lang{\A_\AG}=\AG\lang\A$.
\end{lem}
\begin{proof}
Roughly speaking, the idea is to construct an alternating automaton
that follows one path (for $\EG$) or spawns through all nodes (for
$\AG$) and in each step spawns a new automaton that verifies that $\A$
was a run on the subtree starting at the current node. This
alternation can be removed by a simple powerset construction.

Formally, let $\A$ be given by the state set $Q$, the initial states $I$
and the transition function $\delta$. Define $Q_\AG=\pow Q$,
$I_\AG=\{M\in\pow Q\mid M\cap I\neq\emptyset\}$, and 
$$\delta_\AG(M,\term) = \{ (M_1,\ldots,M_{\ar\term},\ast,\ldots,\ast) \mid
\begin{array}[t]{l}
[\;\forall q\in M\exists
(q_1,\ldots,q_{\ar\term},\ast,\ldots,\ast)\in\delta(\term,q)\\
\qquad q_1\in M_1\land\ldots\land q_{\ar\term}\in M_{\ar\term}\;]\\
\;\land\; \forall i (M_i\cap I\neq\emptyset)\;
\}\;.
\end{array}
$$
Let $\A_\AG$ be the automaton given by this data. Intuitively, the
first condition in the transition function ensures that every state in $M$ can be continued to a
run of $\A$, whereas the second condition ensures that a new run of
$\A$ can be started at every node.

To verify these properties first assume that $t\in\AG\lang\A$. For
every node $p\in t$ set $M_p=\{ q\in Q\mid \run\infty q{t\restrict
p}\}$. Then the mapping $p\mapsto M_p$ is a run of $\A_\AG$ on $t$.
The first condition in the transition relation is fulfilled since
every state that has a infinite run must be able to make a transition
to new states that have an infinite run on the corresponding
subtrees. The second condition is satisfied since $t\in\AG\lang\A$
guarantees that $\A$ has a run for every subtree; so at every subtree,
some initial state has to have a run.

Now assume $t_0\in\lang{\A_\AG}$. So there is a run $r$ of $\A_\AG$ on
$t_0$. We have to show that $t_0\in\AG\lang\A$. To do so, we show that for
\emph{all} trees $t$, 
\emph{all} $M\in Q_\AG$,
if there is \emph{any} run 
of $\A_\AG$ on $t$ starting in $M$ then
for \emph{all} $n\in\NN$, it holds that 
$\forall q\in M.\,\run{n}q{t\restrict p}$.

This indeed shows $t\in\AG\lang\A$. By the properties of $I_\AG$ and
$\delta_\AG$ we immediately get that for all $p\in t_0$ the set $r(p)$
contains an element $q_p\in I$. Applying the claim to $t_0\restrict p$
we obtain that $\A$ has a run, starting in $q_p$ on $t\restrict p$.

So let us show the claim. We argue by induction on $n$. For $n=0$
there's nothing to show. So let $n\geq 1$ and $q\in M$. Assume that
$t$ is of the form $t=\term t_1\ldots t_{\ar\term}$
and let $M_1,\ldots,M_{\ar\term}$ the
states of the run of $\A_\AG$ at the children the root. Since
$(M_1,\ldots,M_{\ar\term},\ldots)\in\delta_\AG(M,\term)$ there exist
$q_1,\ldots,q_{\ar\term}$ such that
$(q_1,\ldots,q_{\ar\term},\ldots)\in\delta(q,\term)$ and $q_i\in
M_i$. Applying the induction hypothesis to $M_i$ and $t_i$ we get
$\run{n-1} {q_i} {t_i}$. Together with the transition
$q\mapsto(q_1,\ldots,q_{\term})$ we get $\run nqt$.

The construction for $\A_\EG$ is similar.
\end{proof}

Taking stock, we see that quite a few safety properties can be
expressed by trivial
automata. Proposition~\ref{prop:trivial-autom:letter} and
Lemmata~\ref{lem:trivial-autom:disj}, \ref{lem:trivial-autom:conj},
\ref{lem:trivial-autom:next}, and~\ref{lem:trivial-autom:gloabally}
show that the fragment of $\mathsf{CTL}$ given by the following
grammar can be expressed by trivial automata.
$$\varphi,\psi\BNFdef \term\BNFor\varphi\lor\psi\BNFor\varphi\land\psi\BNFor
\EX\varphi\BNFor\AX\varphi\BNFor\EG\varphi\BNFor\AG\varphi$$
Of course $\neg\term$ can be expressed by an appropriate disjunction
over all the other letters of the alphabet.

Even though this grammar probably does not exhaust all the properties
expressible by trivial automata, it gives the right flair of the properties being
safety properties. We will now show that the simplest liveness
property, that is the ``eventually'' modality, cannot be expressed, not
even for word languages.

\begin{DEF}[Word Alphabet]
An alphabet $\Sigma$ is called a \emph{word alphabet}, if all its
letters $\term\in\Sigma$ have arity $\ar\term=1$.
\end{DEF}

\begin{REM}
If $\Sigma$ is a word alphabet, then the only $\Sigma$-terms are
$\omega$-words.
\end{REM}

\begin{lem}[Pumping Lemma for Trivial Automata over
	Words]\label{lem:pumping-trivial-automata}
Let $\A$ be a trivial automaton over a word alphabet $\Sigma$. Then
there is a natural number $n$ such that for every word $w$ such that
$\runIn n w$ there is a prefix of $w$ of the form $uv$ with $|uv|\leq
n$ and $|v|\geq 1$ such that $uv^\omega\in\lang\A$.
\end{lem}
\begin{proof}
Set $n=|Q|+1$ where $Q$ is the set of states of $\A$. Let
$w=\term_0\term_1\term_2\ldots$ and assume $\runIn n w$. Let the
states $q_0 q_1 \ldots q_{n-1}$ constitute such a run up to level $n$
on $w$. Since $|Q|=n-1$ there must be $0\leq i<j<n$ such that
$q_i=q_j$. Set $u=\term_0\ldots\term_{i-1}$ and
$v=\term_i\ldots\term_{j-1}$. Then $q_0 \ldots q_{i-1}(q_i\ldots
q_{j-1})^\omega$ constitutes a run on $uv^\omega$ and $u$, $v$ are as
desired.
\end{proof}

An immediate consequence is, that trivial automata cannot express the
property ``eventually $b$'', as the following corollary shows.

\begin{cor}
The language $\ll=a^\ast b(a+b)^\omega$ is not the language of any trivial
automaton.
\end{cor}
\begin{proof}
Suppose, for sake of contradiction, that $\ll=\lang\A$ for some
trivial automaton $\A$ and let $n$ be as asserted by
Lemma~\ref{lem:pumping-trivial-automata}. Consider
$a^nba^\omega\in\ll=\lang\A$ and let $u$, $v$ be as asserted by the
lemma. Since $uv$ is a prefix of $a^nba^\omega$ of length at most $n$,
both, $u$ and $v$ must consist of letters $a$ only, and therefore the lemma
asserts $a^\omega\in\lang\A=\ll$ which is not the case. 
\end{proof}
\end{section}

\section{Infinitary Lambda Trees}\label{sEc:inf-lam-trees}
Now let $\Sigma'$ be a fixed set of letters and let $\term$ from now
on only range over elements of $\Sigma'$. The choice of the name
$\Sigma'$ will become clear in
Definition~\ref{def:full-alphabet-sigma}, when we have to extend the
alphabet in the context of continuous normalisation.

\begin{DEF}
The {\em simple types}, denoted by $\rho$, $\sigma$, $\tau$, are built
from the base type $\iota$ by arrows $\rho\to\sigma$. The arrow
associates to the right. In particular, $\vec\rho\to\iota$ is short
for $\rho_1\to(\rho_2\to(\ldots(\rho_n\to\iota)\ldots))$.
\end{DEF}

In the lambda calculus the most common way to from terms is via
application.
In lambda-trees application is represented by a binary $\app$-node. In
linear notation, we omit the ``$\app$'' and write a tree consisting of
an $\app$-node at the root and subtrees $s$ and $t$ just as
juxtaposition $st$. Application associates to the right, i.e., $rst$
is short for $((rs)t)$. 
\begin{DEF}
The
{\em infinitary simply-typed lambda-trees} over typed terminals
$\Sigma'$ are coinductively given by the grammar
$$r,s\BNFdef x^\rho\BNFor (\lambda x^\rho t^\sigma)^{\rho\to\sigma}
\BNFor (t^{\rho\to\sigma}s^\rho)^\sigma\BNFor
 \term^{\iota\to\ldots\to\iota\to\iota}\,.
$$
In other words, they are,
not-necessarily well founded, trees built, in a 
locally type respecting way,  from
unary $\lambda x^\rho$-nodes, binary $\app$-nodes representing
application, and leaf nodes consisting of
typed variables $x^\rho$ of type $\rho$ and
typed constants $\term\in\Sigma'$ of type
  $\underbrace{\iota\to\ldots\to\iota}_{\ar\term}\to\iota$.

Here $\lambda x^\rho$ binds free occurrences of the variable $x^\rho$
in its body. Trees with all variables bound are called {\em closed}.

A lambda-tree with only finitely many non-isomorphic subtrees is
called \emph{regular}.
\end{DEF}

We omit type superscripts if they are clear from the context, or
irrelevant.

We usually leave out the words ``simply typed'', tacitly assuming all our
lambda-trees to be simply typed and to use terminals from $\Sigma'$
only. 
Figure~\ref{fig:lambdatree} shows two regular lambda-trees. Arrows are
used to show where the pattern repeats, or to draw isomorphic subtrees
only once. Note that they denote terms (shown in
Figure~\ref{fig:valuetree}) that are not regular. Here, by ``denote''
we mean the \emph{term reading} of the normal form.
 
\begin{figure}
\centerline{%
{%
\setlength\unitlength{12pt}
\begin{picture}(11,11)(-5,-6.5)
\put (-0.5,4) {\makebox[1\unitlength]{$\app$}}
\put (0,4) {\line(1,-1) 1}
\put (0,4){\line(-1,-1) 1}
\put (0.5,2.3) {\makebox[1\unitlength]{$\mathtt a$}}
\put (-1.5,2.3) {\makebox[1\unitlength]{$\lambda x$}}
\put (-1,2) {\line(0,-1) 1}
\put (-1.5,0.3) {\makebox[1\unitlength]{$\app$}}
\put (-1,0) {\line(3,-1) 3}
\put (-1,0){\line(-1,-1) 1}
\put (-2.5,-1.7) {\makebox[1\unitlength]{$\app$}}
\put (-2,-2) {\line(1,-1) 1}
\put (-2,-2) {\line(-1,-1) 1}
\put (-3.5,-3.7) {\makebox[1\unitlength]{$\mathtt f$}}
\put (-1.5,-3.7) {\makebox[1\unitlength]{$x$}}
\put (1.5,-1.7) {\makebox[1\unitlength]{$\app$}}
\put (2,-2) {\line(2,-1) 2}
\put (3.5,-3.7) {\makebox[1\unitlength]{$\app$}}
\put (4,-4) {\line(-1,-1) 1}
\put (4,-4) {\line(1,-1) 1}
\put (2.5,-5.7) {\makebox[1\unitlength]{$\mathtt g$}}
\put (4.5,-5.7) {\makebox[1\unitlength]{$x$}}
\put (-3,1.5) {\vector(2,1) 1}
\qbezier(2,-2)(-5,-9)(-5,-5)
\qbezier(-5,-5)(-5,0.5)(-3,1.5)
\end{picture}%
}%
\qquad%
{%
\setlength\unitlength{12pt}
\begin{picture}(11,20)(-4,-15)
\put (-0.5,4) {\makebox[1\unitlength]{$\app$}}
\put (0,4){\line(-1,-1) 1}
\put (0,4) {\line(3,-1) 3}
\put (2.5,2.3) {\makebox[1\unitlength]{$\app$}}
\put (3,2) {\line(1,-1) 1}
\put (3.5,0.3) {\makebox[1\unitlength]{$\mathtt g$}}
\put (-1.5,2.3) {\makebox[1\unitlength]{$\lambda\varphi$}}
\put (-1,2) {\line(0,-1) 1}
\put (-1.5,0.3) {\makebox[1\unitlength]{$\app$}}
\put (-1,0) {\line(3,-1) 3}
\put (-1,0){\line(-1,-1) 1}
\put (-2.5,-1.7) {\makebox[1\unitlength]{$\app$}}
\put (-2,-2) {\line(1,-1) 1}
\put (-2,-2){\line(-1,-1) 1}
\put (-3.5,-3.7) {\makebox[1\unitlength]{$\mathtt f$}}
\put (1.5,-1.7) {\makebox[1\unitlength]{$\app$}}
\put (2,-2) {\line(2,-1) 2}
\put (3.5,-3.7) {\makebox[1\unitlength]{$\app$}}
\put (4,-4) {\line(-1,-1) 1}
\put (4,-4) {\line(1,-1) 1}
\put (4.5,-5.7) {\makebox[1\unitlength]{$\varphi$}}
%
%
\put (-1.5,-3.7) {\makebox[1\unitlength]{$\app$}}
\put (-1,-4) {\line(1,-1) 1}
\put (-1,-4) {\line(-1,-1) 1}
\put (-2.5,-5.7) {\makebox[1\unitlength]{$\varphi$}}
\put (-0.5,-5.7) {\makebox[1\unitlength]{$\mathtt a$}}
\put (2.5,-5.7) {\makebox[1\unitlength]{$\lambda\varphi$}}
\put (3,-6) {\line(0,-1) 1}
\put (2.5,-7.7) {\makebox[1\unitlength]{$\lambda x$}}
\put (3,-8) {\line(0,-1) 1}
\put (2.5,-9.7) {\makebox[1\unitlength]{$\app$}}
\put (3,-10) {\line(-1,-1) 1}
\put (3,-10) {\line(1,-1) 1}
\put (1.5,-11.7) {\makebox[1\unitlength]{$\varphi$}}
\put (3.5,-11.7) {\makebox[1\unitlength]{$\app$}}
\put (4,-12) {\line(-1,-1) 1}
\put (4,-12) {\line(1,-1) 1}
\put (2.5,-13.7) {\makebox[1\unitlength]{$\varphi$}}
\put (4.5,-13.7) {\makebox[1\unitlength]{$x$}}
%
\qbezier(2,-2)(1,-3)(1,-6)
\qbezier(1,-6)(1,-8)(-1.5,-8)
\qbezier(-1.5,-8)(-4,-8)(-4,-3)
\qbezier(-4,-3)(-4,1)(-3,1.5)
\put (-3,1.5) {\vector(2,1) 1}
\qbezier(3,2)(1,0)(3,-1)
\qbezier(3,-1)(7,-3)(7,-6)
\qbezier(7,-6)(7,-9)(5,-7)
\put (5,-7){\vector(-1,1) 1}
\end{picture}%
}%
}

\caption{Two regular lambda-trees with denotation being the $\{\mathtt
  f,\mathtt g,\mathtt a\}$-terms in
  Figure~\ref{fig:valuetree}.\label{fig:lambdatree}}
\end{figure}

\begin{REM}
It should be noted that in lambda-trees, as opposed to
$\Sigma'$-terms, all constants and variables, no matter what their type
is, occur at leaf positions.

The reason is, that in a lambda-calculus setting the main concept is
that of an application. This is different from first order terms,
where the constructors are the main concept.
Note that we use lambda-trees to denote $\Sigma'$-terms. As these are
different concepts, even normal lambda-trees differ from their
denotation. For example the lambda-tree
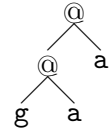
\raisebox{-32pt}{%
{
\setlength\unitlength{10pt}
\begin{picture}(3.4,4.5)(-2,-4)
\put (-0.5,0) {\makebox[1\unitlength]{$\app$}}
\put (0,0) {\line(-1,-1) 1}
\put (0,0) {\line(1,-1) 1}
\put (-1.5,-2) {\makebox[1\unitlength]{$\app$}}
\put (-1,-2) {\line(-1,-1) 1}
\put (-1,-2) {\line(1,-1) 1}
\put (0.5,-1.7) {\makebox[1\unitlength]{$\mathtt a$}}
\put (-2.5,-3.7) {\makebox[1\unitlength]{$\mathtt g$}}
\put (-0.5,-3.7) {\makebox[1\unitlength]{$\mathtt a$}}
\end{picture}%
}%
} denotes the $\Sigma'$-term
\raisebox{-12pt}{%
{
\setlength\unitlength{10pt}
\begin{picture}(2.8,2.5)(-1.4,-1.7)
\put (-0.5,0.3) {\makebox[1\unitlength]{$\mathtt g$}}
\put (0,0) {\line(-1,-1) 1}
\put (0,0) {\line(1,-1) 1}
\put (-1.5,-1.7) {\makebox[1\unitlength]{$\mathtt a$}}
\put (0.5,-1.7) {\makebox[1\unitlength]{$\mathtt a$}}
\end{picture}%
}%
}.
\end{REM}

\section{Recursion Schemes as Means to Define Regular Lambda
  Trees}\label{sEc:rec-scheme-def}

The interest in infinitary lambda-trees in the verification community
recently arose by the study of recursion schemes. It could be
shown~\cite{KnapikNiwinskiUrzyczyn01,KnapikNiwinskiUrzyczyn02} that
under a certain ``safety'' condition the (infinite) terms generated by recursion
schemes have decidable monadic second order theory. For our purpose it
is enough to consider recursion schemes as a convenient means to define
regular lambda-trees.

\begin{DEF}
\emph{Recursion schemes} are given by a set of first-order terminal symbols,
simply-typed non-terminal symbols and for every non-terminal $F$ an
equation
$$F\Myvec x = e$$
where $e$ is an expression of ground type built up from terminals, non-terminals and
the variables $\Myvec x$ by type-respecting application. There is a
distinguished non-terminal symbol $S$ of ground type, called the
\emph{start symbol}.
\end{DEF}

\begin{DEF}
Each recursion scheme \emph{denotes}, in the obvious way, a partial, in general infinite,
term built from the terminals. Starting from the
start symbol, recursively replace the outer-most non-terminals by their
definitions with the arguments substituted in appropriately.
\end{DEF}

\begin{DEF}
To every recursion scheme is \emph{associated} a regular lambda-tree in
the following way. First replace all equations $F\Myvec x=e$ by
$$F = \lambda\Myvec x.e$$
where the right hand side is read as a lambda term.

Then, starting from the start symbol,
recursively replace
all non-terminals by their definition \emph{without
performing any computations}.
\end{DEF}

\begin{REM} Immediately from the definition we note that 
the $\beta$-normal form of the lambda-tree associated with a recursion
scheme, when read a term, \emph{is} the term denoted by that recursion
scheme.
\end{REM}

\begin{figure}
$$\begin{array}{lcl}
S &=& F \mathtt a \\
Fx &=& \mathtt f x (F (\mathtt g x))
\end{array}
\qquad\qquad
\begin{array}{lcl}
S' &=& F'(W\mathtt g)\\
F'\varphi &=& \mathtt f (\varphi a)(F'(W\varphi))\\
W\varphi x &=& \varphi(\varphi x)
\end{array}
$$

\caption{Two recursion schemes.\label{fig:recschemes}}
\end{figure}

\begin{expl}
Figure~\ref{fig:recschemes} shows two recursion schemes with
non-terminals $F\colon\iota\to\iota$,
$F'\colon(\iota\to\iota)\to\iota$,
$W\colon(\iota\to\iota)\to\iota\to\iota$,
and $S,S'\colon\iota$. Their corresponding lambda-trees are the ones
shown in Figure~\ref{fig:lambdatree}. The sharing of an isomorphic
sub-tree arises as both are translations of the same non-terminal
$W$. As already observed, these recursion schemes denote the terms
shown in Figure~\ref{fig:valuetree}.
\end{expl}

\begin{REM}
{\sloppy
The notion of a recursion scheme wouldn't change if we allowed
$\lambda$-abstractions on the right hand side of the equations; we can
always build the closure and ``factor it out'' as a new
non-terminal. For example, the $W\varphi$ in the definition of $F'$
in Figure~\ref{fig:recschemes}
should be thought of as the factored-out closure 
$(\lambda x.\varphi(\varphi x))$ which is part of a line that
originally looked
$$F'\varphi = \mathtt f (\varphi a)(F'(\lambda x.\varphi(\varphi
x)))\;.$$
}
\end{REM}

\section{Continuous Normalisation for the Lambda Calculus}\label{sEc:cn}

As mentioned in the introduction, we are interested in the question,
whether an automaton $\A$ has a run on the normal form of some lambda-tree $t$. Our
plan to investigate this question is by analysing the term $t$.

However, there is no bound on the number of nodes of $t$ that have to
be inspected, and no bound on the number of beta-reductions to be
carried out, before the first symbol of the normal form is
determined\,---\,if it ever will be. In fact, it may well be that an
infinite simply-typed lambda-tree leaves the normal form undefined at
some point.

\begin{expl}\label{expl:Y-I-diverges}
It should be noted that the typing discipline does not prevent the
problem of undefinedness. This is due to inherently infinitary nature
of recursion schemes. Let $Y\colon(\iota\to\iota)\to\iota$,
$I\colon\iota\to\iota$, and $S\colon\iota$ be non-terminal symbols and
consider the recursion scheme
$$
\begin{array}{lcl}
Y \varphi &=& \varphi (Y \varphi)\\
I x       &=& x\\
S         &=& Y I
\end{array}
$$
with start symbol $S$.

Computing the normal form of the associated lambda-tree gives the
following infinite reduction sequence $S=Y I\red I(Y I)\red Y
I\red\ldots$. Of course, the fact that the computation will never
produce a terminal symbol can, in this example, also be trivially seen from
the fact that the whole recursion scheme does not contain any terminal
symbol.
\end{expl}

Whereas the unboundedness of the number of symbols to be inspected is
merely a huge inconvenience, the possibility of undefinedness makes it
unclear what it even is supposed to mean that ``$\A$ has a run on the
normal form of $t$''\,---\,if there is no such normal form. 

This problem of possible undefinedness of the normal form is similar
to a situation in proof theory, where only strong principles guarantee
the termination of the cut-elimination procedure, whereas the
operation itself can be defined in primitive recursive
arithmetic. Continuous Normalisation was introduced by
Mints~\cite{KreiselMintsSimpson75,Mints75} in order to separate
cut-elimination for semiformal systems from their ordinal analysis.
The operational aspects of normalisation, i.e., the manipulations on
infinitary derivations, are isolated and described independently of
the system's proof theoretic complexity, but at the expense of
introducing the void logical rule
$$\AxiomC{$\Gamma$}
\LeftLabel{$(\rep)$}
\UnaryInfC{$\Gamma$}
\DisplayProof$$
of repetition. Note that this rule is both, logically valid and has
the subformula property.

Using the repetition rule, the cut-elimination operator becomes
primitive recursive and can be studied in its own right. As Mints
observed, this cut-elimination operator can also be applied to
non-wellfounded derivations, resulting in a continuous function on
derivation trees (a concise exposition can be found in an
article~\cite{Buchholz91} by Buchholz).

The possibility to handle infinite computations is particularly
natural in the realm of the lambda calculus, where non-termination
actually does happen. Let us explain the idea of continuous
normalisation for the lambda-calculus~\cite{AehligJoachimski02,AehligJoachimski03}
by considering the recursion scheme in
Example~\ref{expl:Y-I-diverges}. The associated lambda tree is shown
in Figure~\ref{fig:Y-I}. 

\begin{figure}
{%
\setlength\unitlength{12pt}
\begin{picture}(14,12)(-9,-10)
\put (-0.5,0.2) {\makebox[1\unitlength]{$\app$}}
\put (0,0) {\line(-2,-1) 4}
\put (0,0) {\line(2,-1) 4}
\put (3.5,-2.7) {\makebox[1\unitlength]{$\lambda\varx$}}
\put (4,-3) {\line(0,-1) 1}
\put (3.5,-4.7) {\makebox[1\unitlength]{$\varx$}}
\put (-4.5,-2.7){\makebox[1\unitlength]{$\lambda\varphi$}}
\put (-4,-3) {\line(0,-1) 1}
\put (-4.5,-4.8) {\makebox[1\unitlength]{$\app$}}
\put (-4,-5) {\line(2,-1) 2}
\put (-4,-5) {\line(-2,-1) 2}
\put (-6.5,-6.7) {\makebox[1\unitlength]{$\varphi$}}
\put (-2.5,-6.8) {\makebox[1\unitlength]{$\app$}}
\put (-2,-7) {\line (2,-1) 2}
\put (-0.5,-8.7) {\makebox[1\unitlength]{$\varphi$}}
\put (-2,-7) {\line (-2,-1) 2}
\put (-6,-3) {\vector(3,1) {1.3}}
\qbezier(-4,-8)(-6,-9)(-7,-9)
\qbezier(-7,-9)(-9,-9)(-9,-6)
\qbezier(-9,-6)(-9,-4)(-6,-3)
\end{picture}%
}
\caption{The lambda-tree associated to the recursion scheme in
Example~\ref{expl:Y-I-diverges}.\label{fig:Y-I}}
\end{figure}

We look at the outer-most constructor of the term and see an
application. Just from this knowledge we cannot deduce any constructor
of the normal form. The normal form \emph{read as a lambda-tree}
could be an application as well, e.g., if
the left term is a terminal; since we're trying to compute the
normal form as a $\Sigma'$-tree, even in this case we would have to
inspect the term further to find out which terminal it is, the term
starts with. But, more importantly, it could also be that the left term
is a $\lambda$-abstraction, in which a beta-reduction has to be
carried out and the normal form could look almost arbitrary. 
So we don't know any constructor of the normal form yet. On the other hand,
we want to be uniformly continuous with identity as modulus of
continuity; in other words, we want to ensure that the output of all
nodes of level $k$ only depend on the input of level $k$. We solve
this problem by outputting $\rep$, signalling that we have to read
more input to decide what the normal form will look like.

Having output $\rep$ we now may look at the next level of the
term. Seeing the $\lambda\varphi$ we still don't any constructor of
the normal form, but at least we know that we have to wait for a
different reason\,---\,we have to carry out some
computation. Therefore we output a $\beta$ constructor, signalling
that the delay in the output is due to a beta-reduction being
carried out. Note that in a certain sense (made precise in Lemma~\ref{lem:cn:justification})
this $\beta$ ``justifies'' the first
$\rep$-constructor. The application we have seen in the first step
has disappeared due to the beta-reduction being carried out. A
different form of justification would be outputting a $\Sigma'$-term,
where the lambda-tree reading contains an application.
For example the term $\mathtt{fa}$ with $\mathtt f$
and $\mathtt a$ both terminals would have continuous normal form 
$\rep(\mathtt f(\mathtt a))$, with the
$\rep$ justified by the fact that $\mathtt f$ is applied to one
argument $\mathtt a$.

After this beta-reduction the term $I(YI)$ is remaining, so we're
looking at an application again, and, as before, wait by saying
$\rep$. Again, there is a lambda abstraction to the left of the
application, so we say $\beta$ and carry out the reduction due to the
$\lambda x$, leaving us with $YI$, which happens to be the term we
started with. Of course, we don't know this yet, as the only thing we
see so far is the outermost $\app$. But the fact that we arrived at
$YI$ again ensures that the pattern $\rep\beta\rep\beta\ldots$ of the
normal form will repeat.

Let us now formally introduce continuous normalisation. As mentioned,
we extend the language by two new terminals. The $\rep$-constructor
for a delay due to inspection of an application and the
$\beta$-constructor for a delay due to a beta-reduction.

\begin{DEF}\label{def:full-alphabet-sigma}
Define $\Sigma=\Sigma'\cup\{\rep,\beta\}$ with $\rep,\beta$ two new
terminals of arity one.
\end{DEF}

The continuous normalisation procedure, which will compute the
continuous normal form, follows the informal description above. In
other words, if we see an application we output $\rep$ and carry on by
reading more input. If we see a lambda-abstraction our typing
restrictions force that we have to have collected some arguments
before, so that a beta-reduction has to be carried out, accompanied by
a $\beta$ constructor; in the more general case~\cite{AehligJoachimski02} of the untyped lambda
calculus~\cite{Barendregt77} we would have to do a case distinction on
whether we have at least one argument collected or not. In the latter
case the normal form would start with a $\lambda$. Finally, if we find
a terminal symbol we construct a term, which is the terminal symbol
applied to the continuous normal forms of the arguments collected so
far.

In our official Definition~\ref{def:cn} of the continuous normal form,
the expression $\CN t{\vec t}$ should be read as ``the continuous
normal form of $t$, with arguments $t_1,\ldots, t_n$ collected
already''. Correspondingly the continuous normal form of $t$ is $\CN t
{()}$ which we also abbreviate by $t^\beta$.

\begin{DEF}\label{def:cn}
For $t$, $\Myvec t$ closed infinitary simply-typed lambda-trees such
that $t\Myvec t$ is of ground type we define a $\Sigma$-term
$\CN t{\Myvec t}$ coinductively as follows.
$$\begin{array}{ll}
\CN{(rs)}{\Myvec t} &=\rep(\CN r{(s,\Myvec t)})\\
\CN{(\lambda x.r)}{(s,\Myvec t)} &=\beta (\CN{r\subst s x}{\Myvec t})\\
\CN\term{\Myvec t} &= \term(t_1^\beta,\ldots,t_n^\beta)
\end{array}$$
Here we used $r\subst s x$ to denote the substitution of $s$ for $x$ in
$r$. This substitution is necessarily capture free as $s$ is
closed. By $\term(T_1,\ldots,T_n)$ we denote the term with label
$\term$ at the root and $T_1,\ldots,T_n$ as its $n$ children; this
includes the case $n=0$, where $\term()$ denotes the term consisting
of a single node $\term$. Similar notation is used for $\rep(T)$ and
$\beta(T)$. Moreover we used $r^\beta$ as a shorthand for $\CN r
{()}$.

The term $t^\beta$ is also called the \emph{continuous normal form} of
$t$.
\end{DEF}

A first observation is that the definition obeys the informal idea of
``justifying'' the delay constructors. We note that, whenever the
number of collected arguments increases we output a $\rep$, and
whenever the number of arguments decreases (due to an argument being
consumed by a beta-reduction) we output a $\beta$. This bookkeeping of
the number of collected arguments is made precise in the next lemma.

\begin{lem}\label{lem:cn:justification}
If $\CN
t{(t_1,\ldots,t_k)}=\Cdelay_1(\Cdelay_2(\ldots(\Cdelay_\ell.\term(\vec
s))))$
with $\Cdelay_1,\ldots,\Cdelay_\ell\in\{\rep,\beta\}$ then
the equation
$k+|\{i\mid \Cdelay_i=\rep\}| = |\{i\mid\Cdelay_i=\beta\}| +
\ar\term$ holds.
\end{lem}
\begin{proof}
A simple induction on $\ell$. If $\ell=0$, the claim $k=\ar\term$ follows
from the typing requirements. Note that we allowed the expression $\CN
t{\vec t}$ only of $t\vec t$ is well typed of ground type. If $\ell>0$ we
distinguish whether $t$ is an application or a lambda-abstraction. In
either case we unfold the definition of $\CN t{\vec t}$ once and can
apply the induction hypothesis.
\end{proof}

Next we will study the relation between lambda terms, their continuous
normal forms, and their normal forms in the usual sense, in case the
latter exists. This, on the one hand, will give a clearer picture on
what the continuous normal form of a lambda term is. On the other
hand, it will also justify the claim, that is not only technically
more convenient for the development in the rest of this
article to use continuous normalisation, but that it is also more
informative.

As an immediate observation, the reader might note that any property
expressible by some automaton $\A$ working on $\Sigma'$-trees can be
lifted to a property on $\Sigma$-trees by ``ignoring the additional
$\rep$ and $\beta$ constructors''. The lifted property can also be
expressed by an automaton. We just have to extend the transition
function $\delta$ by setting
$\delta(q,\rep)=\delta(q,\beta)=\{(q,\ast,\ldots,\ast)\}$. In
particular, using continuous normalisation does not cause any
disadvantages for the decision problem we are interested in.

We already mentioned that output up to depth $h$ only depends on the input up to
depth $h$. To make this idea precise, we first define a notion of
similarity for lambda-tree or $\Sigma$-terms. The relation $r\Tsim k
s$ holds, if $r$ and $s$ coincide up to level $k$. This is made
precise in the following definition.

\begin{DEF}
For $\Sigma$-terms $r$, $s$ we define, by induction on $k$, the
relation $r\Tsim k s$ by the following rules.
$$
\begin{array}{cc}
\AxiomC{}
\UnaryInfC{$r\Tsim 0 s$}
\DisplayProof
&
\AxiomC{$r_1\Tsim k s_1$, \ldots, $r_\ell \Tsim k s_\ell$}
\UnaryInfC{$\term(r_1,\ldots,r_k)\Tsim{k+1}\term(s_1,\ldots,s_\ell)$}
\DisplayProof
\end{array}
$$

For lambda-trees $r$, $s$ we define, by induction on $k$, the
relation $r\Tsim k s$ by the following rules.
$$
\begin{array}{ccc}
\AxiomC{}
\UnaryInfC{$r\Tsim 0 s$}
\DisplayProof
&
\AxiomC{$r\Tsim k s$}
\UnaryInfC{$\lambda x. r\Tsim{k+1} \lambda x. s$}
\DisplayProof
&
\AxiomC{$r\Tsim k r'$}\AxiomC{$s\Tsim k s'$}
\BinaryInfC{$rs\Tsim{k+1}r's'$}
\DisplayProof
\\
&
\AxiomC{\strut}
\UnaryInfC{$x\Tsim k x$}
\DisplayProof
&
\AxiomC{\strut}
\UnaryInfC{$\term\Tsim k\term$}
\DisplayProof
\end{array}
$$
\end{DEF}

\begin{prop}\label{prop:simrel-weak}
 If $r$ and $s$ are both $\Sigma$-terms or both
lambda-trees and $\ell,k\in\NN$, then $r\Tsim \ell s$ and $\ell\geq k$
imply $r\Tsim k s$.
\end{prop}
\begin{proof}
Induction on $k$.
\end{proof}

\begin{REM}\label{prop:simrel-ultrametric}
Obviously, $s=t$ holds if and only if $\forall k. s\Tsim k
t$. Moreover, each of the relations $\Tsim k$ is an equivalence
relation.
\end{REM}

Proposition~\ref{prop:simrel-weak} and
Remark~\ref{prop:simrel-ultrametric} together show, that we obtain a
metric $d$ if we set $d(s,t)$ to be $0$, if $s=t$ and otherwise set
$d(s,t)=\frac 1 {k+1}$ where $k$ is maximal such that $s\Tsim k t$.
We will now show that continuous normalisation is continuous with
respect to this topology. In fact, we even show a stronger statement
of uniform continuity.

\begin{prop}\label{prop:cn-is-unif-cont}
If $s\Tsim k s'$ and $t_1\Tsim k t_1'$,~\ldots, $t_n\Tsim k t_n'$ then
$\CN s{\vec t}\Tsim k\CN {s'}{\vec t'}$.
\end{prop}
\begin{proof}
Induction on $k$. If $k=0$, there is nothing to show. If $k>0$, then the 
outermost constructors of $s$ and $s'$ have to coincide. We unfold the definitions
of $\CN s{\vec t}$ and $\CN{s'}{\vec t'}$ once and apply the induction hypothesis.
\end{proof}

Now that we know (by Proposition~\ref{prop:cn-is-unif-cont}) that continuous normalisation 
does not consume too much input in order to produce the output, we aim at showing that the
output is actually useful and not just a pointless collection of delay constructors.
We have already seen (in Lemma~\ref{lem:cn:justification}) that
the $\rep$ constructors are justified by either $\beta$ constructors
or the arity of the terminals in the output produced. So what remains to show is, that
the $\beta$ constructors are not arbitrary, but in a reasonable sense related to the 
underlying computation. In fact, it will turn out, that every $\beta$ constructor 
corresponds to a beta reduction in the head normalisation strategy;
compare Lemmata~\ref{lem:cn-nf-implies-red} and~\ref{lem:cn-red-implies-nf}.
It is well known that this reduction strategy finds a normal form, if
there is one.

\begin{lem}\label{lem:cn-nf-implies-red}
If $\CN{t}{\vec t}=\Cdelay_1(\ldots(\Cdelay_k(\term(s_1,\ldots,s_{\ar\term}))))$ 
with $\Cdelay_i\in\{\rep,\beta\}$
then there
are lambda-trees $r_1,\ldots,r_{\ar\term}$ such that
\begin{enumerate}[$\bullet$]
\item $t\vec t$ reduces in $n$ head-reduction steps to $\term\vec r$ where $n$ is the number of $\beta$
      constructors, i.e., $n=|\{i\mid \Cdelay_i=\beta\}|$, and
\item for each $i$ it holds that $r_i^\beta=s_i$.
\end{enumerate}
\end{lem}
\begin{proof}
Induction on $k$. If $k=0$, inspection of Definition~\ref{def:cn} of $\CN
t{\vec t}$ shows that it must be the case that $t=\term$. So, in this
case $\CN\term{\vec t}=\term(t_1^\beta,\ldots,t_{\ar\term}^\beta)$ and we can
take $\vec r$ to be $\vec t$.

If $k>0$ and $\Cdelay_1=\beta$ it must be the case that $t=\lambda
x.t'$. 
Then $\CN{(\lambda x.t')}{(t_1, t_2,\ldots,t_\ell)}
=\beta(\CN{(t'\subst{t_1}x)}{(t_2,\ldots,t_\ell)})$. So 
${\CN{(t'\subst{t_1}x)}{(t_2,\ldots,t_\ell)}}=\Cdelay_2(\ldots(\Cdelay_k(\term(s_1,\ldots,s_{\ar\term}))))$
and the induction hypothesis gives us $\vec r$ with $r_i^\beta=s_i$
such that $t\vec t$ reduces in
$n-1$ steps to $\term\vec r$. Since, moreover, in one head reduction step, $t\vec t = (\lambda x.t')t_1
t_2\ldots t_\ell$ reduces to $t'\subst{t_1}xt_2\ldots t_\ell$, this
yields the claim.
If $k>0$ and $\Cdelay_1=\rep$ the claim is immediate from the
induction hypothesis.
\end{proof}

{\sloppy
\begin{lem}\label{lem:cn-red-implies-nf}
If $t\vec t$ reduces by $n$ head reduction steps to $\term r_1 \ldots
r_{\ar\term}$ then
for some $\Cdelay_1,\ldots,\Cdelay_k\in\{\rep,\beta\}$ with 
$|\{i\mid\Cdelay_i=\beta\}|=n$ we have 
$\CN t{\vec t}=\Cdelay_1(\ldots(\Cdelay_k(\term(r_1^\beta,\ldots,r_{\ar\term}^\beta))))$.
\end{lem}
\begin{proof}
Induction on $n$. If $n=0$ then $t\vec t$ must be of the form
$\term\vec r$ and, indeed, $\CN t{\vec t} =\rep(\ldots(\rep(\CN \term
{\vec
  r})))=\rep(\ldots(\rep(\term(r_1^\beta,\ldots,r_{\ar\term}^\beta))))$.

If $n>0$ then $t$ is of the form $(\lambda x s)\vec s$. Writing $\vec t
'$ for $\vec s\vec t$ we note that $\CN t{\vec t}
=\rep(\ldots(\rep(\CN{(\lambda x.s)}{\vec
  t'})))=\rep(\ldots(\rep(\beta(\CN{s\subst
  {t_1'}x}{(t_2',\ldots,t_\ell')}))))$. Since the head reduct of $t\vec
t$ is $s\subst{t_1'}xt_2'\ldots t_\ell'$, the induction hypothesis
yields the claim.
\end{proof}
}

It should be noted that in the special case of $\vec t$ being the
empty list, Lemmata~\ref{lem:cn-nf-implies-red}
and~\ref{lem:cn-red-implies-nf} talk 
about the continuous normal form of $t$. 

%

\section{Finitary Semantics and Proof System}\label{sEc:fin-sem}
Let $\A$ be a fixed nondeterministic tree automaton with state set $Q$
and transition function $\delta\colon Q\times
\Sigma\to\pow{(Q\cup\{\ast\})^N}$.
The main technical idea of this article is to use a finite semantics
for the simple types, describing how $\A$ ``sees'' an object of that
type.

\begin{DEF}
For $\tau$ a simple type we define $\sem\tau$ inductively as follows.
$$\begin{array}{ll}
\sem\iota &=\mathfrak P(Q)\\
\sem{\rho\to\sigma} &={}^{\sem\rho}\sem\sigma
\end{array}$$
In other words, we start with the power set of the state set of $\A$ in
the base case, and use the full set theoretic function space for
arrow-types. 
\end{DEF}

\begin{REM}
Obviously all the $\sem\tau$ are finite sets.
\end{REM}

\begin{expl}\label{expl:swap-function}
Taking $\A$ to be the automaton of Example~\ref{expl:prop}, we have
$\sem\iota=\{\emptyset,\{\qeven\},\{\qodd\},Q\}$ and examples of
elements of
$\sem{\iota\to\iota}$ include the identity function $\id$, as well as the
``swap function'' $\swap$ defined by $\swap(\emptyset)=\emptyset$,
$\swap(Q)=Q$, $\swap(\{\qeven\})=\{\qodd\}$, and
$\swap(\{\qodd\})=\{\qeven\}$.
\end{expl}

\begin{DEF}
$\sem\tau$ is partially ordered as follows.
\begin{enumerate}[$\bullet$]
\item For $R,S\in\sem\iota$ we set $R\less S$ iff $R\subseteq S$.
\item For $f,g\in\sem{\rho\to\sigma}$ we set $f\less g$ iff $\forall
  a\in\sem\rho. fa\less ga$.
\end{enumerate}
\end{DEF}

\begin{REM}
Obviously suprema and infima with respect to $\less$ exist.
\end{REM}

We often need the concept ``continue with $f$ after reading one $\rep$
symbol''. We call this $\rep$-lifting. Similar for $\beta$.

\begin{DEF}
For $f\in\sem{\Myvec\rho\to\iota}$ we define the liftings $\rep(f),
\beta(f)\in\sem{\Myvec\rho\to\iota}$ as follows.
$$
\begin{array}{lcl}
\rep(f)(\Myvec a)  &=&\{q\mid\delta(q,\rep)\cap f\Myvec
a\times\{\ast\}\times\ldots\times\{\ast\}\neq\emptyset\}\\
\beta(f)(\Myvec a) &=&\{q\mid\delta(q,\beta)\cap f\Myvec a\times\{\ast\}\times\ldots\times\{\ast\}\neq\emptyset\}
\end{array}
$$
\end{DEF}

\begin{REM}
If $\A$ is obtained from an automaton working on $\Sigma'$-terms by
setting
$\delta(q,\rep)=\delta(q,\beta)=\{(q,\ast,\ldots,\ast)\}$
then $\rep(f)=\beta(f)=f$ for all $f$.
\end{REM}

Using this finite semantics we can use it to annotate a lambda-tree by
semantical values for its subtrees to show that the denoted term has
good properties with respect to $\A$. We start by an example. 

\begin{figure}
\centerline{%
{%
\setlength\unitlength{12pt}
\begin{picture}(34,22)(-15,-17)
\put (-0.5,4) {\makebox[1\unitlength]{$\app$}}
\put (0,4){\line(-1,-1) 1}
\put (0,4) {\line(3,-1) 3}
\put (2.5,2.3) {\makebox[1\unitlength]{$\app$}}
\put (3,2) {\line(1,-1) 1}
\put (3.5,0.3) {\makebox[1\unitlength]{$\mathtt g$}}
\put (-1.5,2.3) {\makebox[1\unitlength]{$\lambda\varphi$}}
\put (-1,2) {\line(0,-1) 1}
\put (-1.5,0.3) {\makebox[1\unitlength]{$\app$}}
\put (-1,0) {\line(3,-1) 3}
\put (-1,0){\line(-1,-1) 1}
\put (-2.5,-1.7) {\makebox[1\unitlength]{$\app$}}
\put (-2,-2) {\line(1,-1) 1}
\put (-2,-2){\line(-1,-1) 1}
\put (-3.5,-3.7) {\makebox[1\unitlength]{$\mathtt f$}}
\put (1.5,-1.7) {\makebox[1\unitlength]{$\app$}}
\put (2,-2) {\line(2,-1) 2}
\put (3.5,-3.7) {\makebox[1\unitlength]{$\app$}}
\put (4,-4) {\line(-1,-1) 1}
\put (4,-4) {\line(1,-1) 1}
\put (4.5,-5.7) {\makebox[1\unitlength]{$\varphi$}}
%
\put (-1.5,-3.7) {\makebox[1\unitlength]{$\app$}}
\put (-1,-4) {\line(1,-1) 1}
\put (-1,-4) {\line(-1,-1) 1}
\put (-2.5,-5.7) {\makebox[1\unitlength]{$\varphi$}}
\put (-0.5,-5.7) {\makebox[1\unitlength]{$\mathtt a$}}
\put (2.5,-5.7) {\makebox[1\unitlength]{$\lambda\varphi$}}
\put (3,-6) {\line(0,-1) 1}
\put (2.5,-7.7) {\makebox[1\unitlength]{$\lambda x$}}
\put (3,-8) {\line(0,-1) 1}
\put (2.5,-9.7) {\makebox[1\unitlength]{$\app$}}
\put (3,-10) {\line(-1,-1) 1}
\put (3,-10) {\line(1,-1) 1}
\put (1.5,-11.7) {\makebox[1\unitlength]{$\varphi$}}
\put (3.5,-11.7) {\makebox[1\unitlength]{$\app$}}
\put (4,-12) {\line(-1,-1) 1}
\put (4,-12) {\line(1,-1) 1}
\put (2.5,-13.7) {\makebox[1\unitlength]{$\varphi$}}
\put (4.5,-13.7) {\makebox[1\unitlength]{$x$}}
\qbezier(2,-2)(1,-3)(1,-6)
\qbezier(1,-6)(1,-8)(-1.5,-8)
\qbezier(-1.5,-8)(-4,-8)(-4,-3)
\qbezier(-4,-3)(-4,1)(-3,1.5)
\put (-3,1.5) {\vector(2,1) 1}
\qbezier(3,2)(1,0)(3,-1)
\qbezier(3,-1)(7,-3)(7,-6)
\qbezier(7,-6)(7,-9)(5,-7)
\put (5,-7){\vector(-1,1) 1}
\put(-15.5,2.3){\framebox[6\unitlength]{$\id\mapsto\{\qeven\}$}}
\put(-9.5,2.5){\vector(1,0){8}}
\put(-15.5,0.3){\framebox[6\unitlength]{$\Gamma_\varphi\vdash\{\qeven\}$}}
\put(-9.5,0.5){\vector(1,0){8}}
\put(-15.5,-1.7){\framebox[8\unitlength]{$\Gamma_\varphi\vdash\{\qeven\}\mapsto\{\qeven\}$}}
\put(-7.5,-1.5){\vector(1,0)5}
\put(-15.5,-3.7){\framebox[11\unitlength]{$\Gamma_\varphi\vdash\{\qeven\}\mapsto\{\qeven\}
                                           \mapsto\{\qeven\}$}}
\put(-4.5,-3.5){\vector(1,0)1}
\put(-15.5,-5.7){\framebox[6\unitlength]{$\Gamma_\varphi\vdash\{\qeven\}$}}
\put(-9.5,-5.5){\line(1,0){6.5}}\put(-3,-5.5){\vector(1,1){1.5}}
\put(-15.5,-7.7){\framebox[6\unitlength]{$\Gamma_\varphi\vdash\id$}}
\put(-9.5,-7.5){\line(1,0){5.5}}\put(-4,-7.5){\vector(1,1){1.5}}
\put(-15.5,-9.7){\framebox[6\unitlength]{$\Gamma_\varphi\vdash\{\qeven\}$}}
\put(-9.5,-9.5){\line(1,0){5.5}}\put(-4,-9.5){\vector(1,1){3.5}}
\put(-15.5,-11.7){\framebox[10\unitlength]{$\id\mapsto\id\text{~~and~~}\swap\mapsto\id$}}
\put(-5.5,-11.5){\line(1,0){2.5}}\put(-3,-11.5){\vector(1,1){5.5}}
\put(-15.5,-13.7){\framebox[12\unitlength]{$\begin{array}[t]{l}
\Gamma_{\varphi,x}\vdash\id\text{~~and~~}\Gamma'_{\varphi,x}\vdash\swap\\
\Gamma_{\varphi,x'}\vdash\id\text{~~and~~}\Gamma'_{\varphi,x'}\vdash\swap\\
\end{array}
$} }
\put(-3.5,-13.5){\vector(1,0){6}}
\put(0,-13.5){\vector(1,1){1.5}}
\put(8.5,4.5){\fbox{$\{\qeven\}$}}\put(8.5,4.5){\vector(-1,0)8}
\put(8.5,2.5){\fbox{$\id$}}\put(8.5,2.5){\vector(-1,0)5}
\put(8.5,0.5){\fbox{$\swap$}}\put(8.5,0.5){\vector(-1,0)4}
\put(8.5,-1.5){\fbox{$\Gamma_\varphi\vdash\{\qeven\}$}}\put(8.5,-1.5){\vector(-1,0)6}
\put(8.5,-3.5){\fbox{$\Gamma_\varphi\vdash\id$}}\put(8.5,-3.5){\vector(-1,0)4}
\put(8.5,-5.5){\fbox{$\Gamma_\varphi\vdash\id$}}\put(8.5,-5.5){\vector(-1,0)3}
\put(8.5,-7.5){\fbox{$\Gamma_\varphi\vdash\id\text{~~and~~}\Gamma'_\varphi\vdash\id$}}
\put(8.5,-7.5){\vector(-1,0)5}
\put(6.5,-9.5){\fbox{$\begin{array}[t]{l}
\Gamma_{\varphi,x}\vdash\{\qeven\}\text{~~and~~}\Gamma'_{\varphi,x}\vdash\{\qeven\}\\
\Gamma_{\varphi,x'}\vdash\{\qodd\}\text{~~and~~}\Gamma'_{\varphi,x'}\vdash\{\qodd\}\\
\end{array}
$}}
\put(6.5,-9.5){\vector(-1,0)3}
\put(6.5,-12.5){\fbox{$\begin{array}[t]{l}
\Gamma_{\varphi,x}\vdash\{\qeven\}\text{~~and~~}\Gamma'_{\varphi,x}\vdash\{\qodd\}\\
\Gamma_{\varphi,x'}\vdash\{\qodd\}\text{~~and~~}\Gamma'_{\varphi,x'}\vdash\{\qeven\}\\
\end{array}
$}}
\put(6.5,-12.5){\line(-1,0)1}\put(5.5,-12.5){\vector(-1,1)1}
\put(6.5,-15.5){\fbox{$\begin{array}[t]{l}
\Gamma_{\varphi,x}\vdash\{\qeven\}\text{~~and~~}\Gamma'_{\varphi,x}\vdash\{\qeven\}\\
\Gamma_{\varphi,x'}\vdash\{\qodd\}\text{~~and~~}\Gamma'_{\varphi,x'}\vdash\{\qodd\}\\
\end{array}
$}}
\put(6.5,-15.5){\line(-1,0){1.5}}\put(5,-15.5){\vector(0,1){1.5}}
\end{picture}%
}%
}

\caption{A proof that $\A$ has an infinite run starting in $\qeven$ on
  the denoted term.\label{fig:inf-proof}} 
\end{figure}

\begin{expl}\label{expl:intro-to-inf-proof}
The
second recursion scheme in Figure~\ref{fig:recschemes} denotes a term
where the ``side branches'' contain $2,4,8,\ldots,2^n,\ldots$ times
the letter $\mathtt g$. As these are all even numbers, the automaton 
$\A$ of Example~\ref{expl:prop} should
have a run starting in $\qeven$.

We now informally argue how a formal ``proof'' of this fact can be
obtained by assigning semantical values to the nodes
of the corresponding lambda-tree, which is the right tree in
Figure~\ref{fig:lambdatree}. The notion of ``proof'' will be made
formal in Definition~\ref{def:proof-system}.

So we start by assigning the root $\{\qeven\}\in\sem\iota$. Since the
term is an application, we have to guess the semantics of the
argument (of type $\iota\to\iota$). Our (correct) guess is, that it
keeps the parity of $\mathtt g$s unchanged, hence our guess is $\id$;
the function side then must be something that maps $\id$ to
$\{\qeven\}$. Let us denote by $\id\mapsto\{\qeven\}$ the function in
${}^{\sem{\iota\to\iota}}\sem{\iota}$ defined by
$(\id\mapsto\{\qeven\})(\id)=\{\qeven\}$ and
$(\id\mapsto\{\qeven\})(f)=\emptyset$ if $f\neq\id$. 

The next node to the left is an abstraction. So we have to assign the
body the value $\{\qeven\}$ in a context where $\varphi$ is mapped
to $\id$. Let us
denote this context by $\Gamma_\varphi$.

In a similar way we fill out the remaining
annotations. Figure~\ref{fig:inf-proof} shows the whole proof. Here
$\Gamma'_\varphi$ is the context that maps $\varphi$ to $\swap$;
moreover $\Gamma_{\varphi,x}$, $\Gamma'_{\varphi,x}$,
$\Gamma_{\varphi,x'}$, and $\Gamma'_{\varphi,x'}$ are the same as
$\Gamma_\varphi$ and $\Gamma'_\varphi$ but with $x$ mapped to
$\{\qeven\}$ and $\{\qodd\}$, respectively.

It should be noted that a similar attempt to assign semantical values
to the other lambda-tree in Figure~\ref{fig:lambdatree} fails at the
down-most $x$ where in the context $\Gamma$ with
$\Gamma(x)=\{\qeven\}$ we cannot assign $x$ the value $\{\qodd\}$.
\end{expl}

To make the intuition of the example precise, we formally define a
``proof system'' of possible annotations $(\Gamma,a)$ for a
(sub)tree. Since the $\sem\tau$ are all finite sets, there are only
finitely many possible annotations.

To simplify the later argument of our proof, which otherwise would 
be coinductive, we add a level $n$ to our notion of proof. This
level should be interpreted as ``for up to $n$ steps we can pretend to
have a proof''. This reflects the fact that coinduction is nothing but
induction on observations.

\begin{DEF}\label{def:context}
A \emph{context} is a finite mapping from variables $x^\sigma$ to
their corresponding semantics $\sem\sigma$. We use $\Gamma$ to range
over contexts.

If $\Gamma$ is a context, $x$ a variable of type $\sigma$ and
$a\in\sem\sigma$ we denote by $\Gamma_x^a$ the context $\Gamma$
modified in that $x$ is mapped to $a$, regardless of whether $x$ was
or was not in the domain of $\Gamma$.
\end{DEF}

\begin{DEF}\label{def:proof-system}
For $\Gamma$ a context, $a\in\sem\rho$ a value, and $t$ an
infinitary, maybe open, lambda-tree of type $\rho$, with free variables among
$\dom\Gamma$, we define
$$\fproof n \Gamma a t \rho$$
by induction on the natural number $n$ as follows.
\begin{enumerate}[$\bullet$]
\item $\fproof0\Gamma a t\rho$ always holds.
\item $\fproof n\Gamma a {x_i}\rho$ holds, provided
  $a\less\Gamma(x_i)$.
\item $\fproof{n+1}\Gamma a{st}\sigma$ holds, provided
there exists $f\in\sem{\rho\to\sigma}$, $u\in\sem\rho$ such that $a\less\rep(fu)$,
$\fproof n\Gamma fs{\rho\to\sigma}$, and $\fproof n\Gamma u t\rho$.
\item 
$\fproof{n+1}\Gamma f {\lambda x^\rho.s}{\rho\to\sigma}$ holds,
  provided for all $a\in\sem\rho$ there is a $b_a\in\sem\sigma$ such that
      $fa\less\beta(b_a)$ and $\fproof n{\Gamma_x^a}{b_a}s\sigma$.
\item $\fproof n\Gamma f\term{\iota\to\ldots\to\iota\to\iota}$ holds,
  provided for all $\Myvec a\in\sem{\Myvec\iota}$ we have $f\Myvec
  a\subset\{q\mid\delta(q,\term)\cap a_1\times\ldots\times
  a_{\ar\term}\times\{\ast\}\times\ldots\times\{\ast\}\neq\emptyset\}$.
\end{enumerate}
It should be noted that all the quantifiers in the rules range over finite
sets. Hence the correctness of a rule application can be checked
effectively (and even by a finite automaton).

We write $\fproof\infty\Gamma a t\rho$ to denote $\forall n.\fproof
n\Gamma a t\rho$.
\end{DEF}

{\sloppy
\begin{REM}\label{rem-proof-weak}
Obviously $\fproof{n+1}\Gamma a t\rho$ implies $\fproof n\Gamma a t
\rho$. Moreover, $a'\less a$ and $\fproof n\Gamma at\rho$ imply
$\fproof n\Gamma{a'}t\rho$. Finally, $\fproof n{\Gamma} at\rho$, if
$\fproof n{\Gamma'} at\rho$ for some $\Gamma'$ which agrees with
$\Gamma$ on the free variables of $t$.

Also, in the second an in the last clause we may assume without loss of generality,
that $n>0$. However, this assumption is not necessary, and it is even technically more convenient
not to do so.
\end{REM}
}

\begin{REM}
We notice that the proof informally given in Example~\ref{expl:intro-to-inf-proof} 
and shown in Figure~\ref{fig:inf-proof} complies with the formal
Definition~\ref{def:proof-system}. Indeed, the annotations shown in the figure are
valid for any $n$.
\end{REM}

As already mentioned, for $t$ a term with finitely many free
variables,
the annotations $(\Gamma, a)$ come from a fixed finite set, since
we can restrict $\Gamma$ to the set of free variables of $t$. If,
moreover, $t$ has only finitely many different sub-trees, that is to
say, if $t$ is regular, then only finitely many terms $t$ have to be
considered. So we obtain

\begin{prop}\label{prop-proof-dec}
For $t$ regular, it is decidable whether $\fproof\infty\Gamma a t\rho$.
\end{prop}

Before we continue and show our calculus 
to be sound
(Section~\ref{sEc:sound})
and complete (Section~\ref{sEc:complete}) let us step back and
see what we will then have achieved, once our calculus is proven sound
and complete.

Proposition~\ref{prop-proof-dec} gives us decidability for terms
denoted by regular lambda-trees, and hence in particular for trees
obtained by recursion schemes. Moreover, since the annotations only
have to fit locally, individual subtrees of the lambda-tree can be
verified separately. This is of interest, as for each non-terminal a
separate subtree is generated. In other words, this approach allows
for modular verification; think of the different non-terminals as
different subroutines.
As the semantics is the set-theoretic
one, the annotations are clear enough to be meaningful, if we
have chosen our automaton in such a way that the individual states can
be interpreted extensionally, for example as ``even'' versus ``odd''
number of $\mathtt g$s.

It should also be noted, that the number of possible annotations
only depends on the type of the subtree, and on $\A$, that is, the
property to be investigated.
Fixing $\A$ and the allowed types (which both usually tend
to be quite small), the amount of work to be carried out grows only
linearly with the representation of $t$ as a regular lambda-tree. For
every node we have to make a guess and we have to check whether this
guess is consistent with the guesses for the (at most two) child
nodes. Given that the number of nodes of the representation of $t$
grows linearly with the size of the recursion scheme, the problem is
in fixed-parameter-$\mathcal{NP}$, which doesn't seem too bad for
practical applications.

\section{Truth Relation and Proof of Soundness}\label{sEc:sound}

The soundness of a calculus is usually shown by using a logical
relation, that is, a relation indexed by a type that interprets the
type arrow ``$\to$'' as logical arrow ``$\Rightarrow$''; in other
words,
we define partial truth predicates for the individual
types~\cite{Tait67}.

Since we want to do induction on the ``observation depth'' $n$ of our
proof $\fproof n\cdot\cdot\cdot\tau$ we have to include that depth in
the definition of our truth predicates $\force n\cdot\cdot\tau$. For
technical reasons we have to build in weakening on this depth as well.

\begin{DEF}\label{def-force}
For $f\in\sem{\Myvec\rho\to\iota}$, $n\in\NN$, $t$ a closed
infinitary lambda tree 
of type $\Myvec\rho\to\iota$, the relation $\force n f t
{\Myvec\rho\to\iota}$ is defined by induction on the type as follows.
$$\begin{array}{l}
\force n f t {\Myvec\rho\to\iota}\qquad\text{iff}\\
\forall\ell\leq n
\forall\Myvec a\in\sem{\Myvec\rho}\forall\Myvec r:\Myvec\rho\\
\qquad
(\forall i.\; \force \ell {a_i}{r_i}{\rho_i})\Rightarrow\forall q\in f\Myvec
a.\;\run \ell q {\CN t {\Myvec r}}
\end{array}$$
\end{DEF}

\begin{REM}\label{lem-force-weak}
Immediately from the definition we get the following monotonicity
property.

If $f\less f'$ and $\force n {f'} t \rho$ then $\force n f t \rho$.
\end{REM}

\begin{REM}
In the special case $\Myvec\rho=\varepsilon$ we get
$$\force n S t \iota\qquad\text{iff}\qquad
\forall q\in S.\run n q {t^\beta}$$
Here we used that $\forall\ell\leq n.\,\run\ell q s$ iff $\run n q s$. 
\end{REM}

Immediately from the definition we obtain weakening in the level.
\begin{prop}\label{force-down}
If $\force n f t \rho$ then
$\force{n-1}f t\rho$.
\end{prop}

\begin{Th}\label{le-sound}
Assume
$\fproof n\Gamma at\rho$ for some $\Gamma$ with domain
$\{x_1,\ldots,x_2\}$. For all $\ell\leq n$ and all closed terms $\Myvec
t\colon\Myvec\rho$, if
$\forall i.\; \force \ell{\Gamma(x_i)}{t_i}\rho_i$ then
$\force \ell a{t\subst{\Myvec t}{\Myvec x}}\rho$.
\end{Th}
\begin{proof}
{\sloppy
Induction on $n$, cases according to $\fproof n\Gamma at\rho$.
\begin{enumerate}[$\bullet$]
\item Case $\fproof 0\Gamma a t\rho$ always. Use that
  $\force0a\dots\rho$ holds always.

\item Case $\fproof n\Gamma a {x_i}\rho$ because of
  $a\less{\Gamma(x_i)}$.

Assume $\forall i.\,\force \ell{\Gamma(x_i)}{t_i}\rho_i$. We have to show
$\force \ell{a}{\underbrace{x_i\subst{\Myvec t}{\Myvec
	  x}}_{t_i}}\rho$, which follows from one of our assumptions by
Remark~\ref{lem-force-weak}.

\item Case $\fproof{n+1}\Gamma a{st}\sigma$ thanks to 
$f\in\sem{\rho\to\sigma}$, $u\in\sem\rho$ such that $a\less\rep(fu)$,
  $\fproof n\Gamma fs{\rho\to\sigma}$, and 
$\fproof n\Gamma u t\rho$.

Let $\ell\leq n+1$ be given, and $\Myvec t:\Myvec\rho$ such that
$\forall i.\;\force\ell{\Gamma(x_i)}{t_i}{\rho_i}$. We have to show
$\force\ell a{(st)\underbrace{\subst{\Myvec t}{\Myvec x}}_\eta}\sigma$.

Let $\sigma$ have the form $\sigma=\Myvec\sigma\to\iota$.
Let $k\leq\ell$ be given and $\Myvec s:\Myvec \sigma$, 
$c_i\in\sem{\sigma_i}$ such that 
$\force k {c_i}{s_i}{\sigma_i}$. We have to show
for all $q\in a\Myvec c$ that $\run k q{\underbrace{\CN{(s\eta
	  t\eta)}{\Myvec r}}_{\rep.(\CN{s\eta}{(t\eta,\Myvec r)})}}$.

Hence it suffices to show that there is a $\tilde q\in\delta(q,\rep)$
such that $\run{k-1}{\tilde q}{\CN{s\eta}{(t\eta,\Myvec r)}}$.


Since $k\leq\ell\leq n+1$, we have $k-1\leq n$. Using
Proposition~\ref{force-down} various times we obtain $\forall
i.\;\force{k-1}{\Gamma(x_i)}{t_i}{\rho_i}$. Hence we may use the
induction hypotheses to $\fproof n\Gamma fs{\rho\to\sigma}$ and obtain
$\force{k-1}f{s\eta}{\rho\to\sigma}$. Applying the induction to
$\fproof n\Gamma u t\rho$ yields $\force{k-1}u{t\eta}\rho$.

Applying Proposition~\ref{force-down} to 
$\force k {c_i}{s_i}{\sigma_i}$ 
yields
$\force {k-1} {c_i}{s_i}{\sigma_i}$. Therefore
$\forall \hat q\in fu\Myvec c.\;\run{k-1}{\hat
q}{\CN{s\eta}{(t\eta,\Myvec r)}}$.

Since $a\less\rep(fu)$ we get $\forall q\in a\Myvec c\exists\tilde q\in
\delta(q,\rep).\;\tilde q\in fu\Myvec c$. This together with the last
statement yields the claim.

\item
 \emph{Case} $\fproof{n+1}\Gamma f {\lambda x^\rho.s}{\rho\to\sigma}$
thanks to $\forall a\in\sem\rho$ $\exists b_a\in\sem\sigma$ such that
$fa\less\beta(b_a)$ and $\fproof n{\Gamma_x^a}{b_a}s\sigma$. 

Let $\ell\leq n+1$ be given and $\Myvec t:\Myvec\rho$ with $\force\ell 
{\Gamma(x_i)}{t_i}{\rho_i}$.

 We have to show $\force \ell f{(\lambda
  x^\rho s^\sigma)\eta}{\rho\to\sigma}$ where $\eta$ is short for
${\subst{\Myvec t}{\Myvec x}}$. 

Let $\sigma$ have the form $\sigma=\Myvec\sigma\to\iota$.
Let $k\leq\ell$ be given and $r:\rho$, $\Myvec s:\Myvec \sigma$, $c\in\sem\rho$,
$c_i\in\sem{\sigma_i}$ such that 
$\force k c r \rho$, $\force k {c_i}{s_i}{\sigma_i}$. We have to show
for all $q\in fc\Myvec c$ that $\run k q {\underbrace{\CN{(\lambda
	xs)\eta}{(r,\Myvec s)}}_{\beta.\CN{s\eta_x^r}{\Myvec s}}}$.

Hence it suffices to show that there is a $\tilde q\in\delta(q,\beta)$
such that $\run{k-1}{\tilde q}{\CN{s\eta_x^r}{\Myvec s}}$.


We know $\force k c r \rho$; using Proposition~\ref{force-down} we get
$\force{k-1}cr\rho$ and $\forall
i.\;\force{k-1}{\Gamma(x_i)}{t_i}{\rho_i}$. Since $k\leq\ell\leq n+1$
we get $k-1\leq n$, hence we may apply the induction hypothesis to
$\fproof n{\Gamma_x^a}{b_a}s\sigma$ and obtain
$\force{k-1}{b_a}{s\eta_x^r}\sigma$.

Since again by Proposition~\ref{force-down} we also know 
$\force {k-1} {c_i}{s_i}{\sigma_i}$, we obtain for all 
$\hat q\in b_a\Myvec c$ that
$\run{k-1}{\hat q}{\CN{s\eta_x^r}{\Myvec s}}$.

Since $fc\less\beta(b_c)$ we get that $\forall q\in fc\Myvec
c\exists\tilde q\in\delta(q,\beta).\,\tilde q\in b_c\Myvec c$. This,
together with the last statement yields the claim.
\item Case $\fproof n\Gamma f\term{\iota\to\iota}$ thanks to 
$\forall\Myvec a\in\sem\iota.\; f\Myvec a\subset\{q\mid\delta(q,\term)\cap
  \Myvec a\neq\emptyset\}$.

Let $\ell\leq n$ be given and $\Myvec t:\Myvec \rho$ such that
$\forall i.\;\force\ell{\Gamma(x_i)}{t_i}{\rho_i}$. We have to show
$\force\ell f{\underbrace{\term\subst{\Myvec t}{\Myvec x}}_\term}{\iota\to\iota}$.

Let $k\leq\ell$ be given and $\Myvec r:\Myvec \iota$, $\Myvec S\in\sem\iota$ such that
$\force\ell {S_i}{r_i}\iota$. We have to show for all $q\in f\Myvec S$ that
$\run\ell q{\underbrace{\CN\term{\Myvec r}}_{\term \Myvec r^\beta}}$.

From $\force\ell {S_i}{r_i}\iota$ we get $\forall \tilde q_i\in S_i.\;\run\ell
{\tilde q_i}{r_i^\beta}$. Hence the claim follows since 
$\forall q\in f\Myvec S\;\exists\Myvec{\tilde
  q}\in\delta(a,\term).\;\Myvec{\tilde q}\in\Myvec S$.
\qedhere
\end{enumerate}
}
\end{proof}

It should be noted that in the proof of Theorem~\ref{le-sound} in the
cases of the $\lambda$-rule and the application-rule it was possible
to use the induction hypothesis due to the fact that we used
\emph{continuous} normalisation, as opposed to standard normalisation.

\begin{cor}\label{cor-sound}
For $t$ a closed infinitary lambda term we get immediately from
Theorem~\ref{le-sound}
$$\fproof n\emptyset S t \iota\quad\Longrightarrow\quad
\forall q\in S.\;\run n q {t^\beta}$$
In particular, if $\fproof\infty\emptyset St\iota$ then $\forall q\in
S.\;\run\infty q {t^\beta}$.
\end{cor}

\section{The Canonical Semantics and the Proof of
  Completeness}\label{sEc:complete}

If we want to prove that there is an infinite run, then, in the
case of an application $st$, we have to guess a value for the term $t$ ``cut
out''. 

We could assume an actual run be given and analyse the
``communication'', in the sense of game semantics~\cite{HylandOng00},
between the function $s$ and its argument $t$. However, it is simpler
to assign each term a ``canonical semantics'' $\csem t$, roughly
the supremum of all values we have canonical proofs for.

The subscript $\infty$ signifies that we only consider infinite
runs. The reason is that the level $n$ in our proofs $\fproof n \Gamma
a t\rho$ is not a tight bound; whenever we have a proofs of level $n$,
then there are runs for at least $n$ steps, but on the other hand,
runs might be longer than the maximal level of a proof. This is due to
the fact that $\beta$-reduction moves subterms ``downwards'', that is,
further away from the root, and in that way may construct longer
runs. The estimates in our proof calculus, however, have to consider (in
order to be sound) the worst case, that is, that an argument is used
immediately.

Since, in general, the term $t$ may also have free variables, we have
to consider a canonical semantics $\csemC\Gamma t$ with respect to an
environment $\Gamma$.

\begin{DEF}
By induction on the type we define for $t$ a closed infinite
lambda-tree of type $\rho=\Myvec\rho\to\iota$
its canonical semantics $\csem t\in\sem\rho$ as follows.
$$
\csem t (\Myvec a) = \{ q\mid\exists \Myvec s\colon \Myvec\rho\;.\;\;
\csem{\Myvec s}\less\Myvec a\;\land
\run\infty q {\CN t{\Myvec s}}\}
$$
\end{DEF}

\begin{REM}\label{rm-csem-base}
For $t$ a closed term of base type we have $\csem t = \{ q\mid\run\infty
q{t^\beta}\}$.
\end{REM}

\begin{DEF}
For $\Gamma$ a context, $t\colon\rho$ typed in context $\Gamma$ of
type $\rho=\Myvec\rho\to\iota$ we define
$\csemC\Gamma t\in\sem\rho$ by the following explicit definition.
$$
\begin{array}{ll}
\csemC\Gamma t(\Myvec a) = \{ q\mid &
\exists\eta.\;\dom\eta=\dom\Gamma\land\\
&(\forall x\in\dom\Gamma.\eta(x)\text{~closed}\land\csem{\eta(x)}\less\Gamma(x))\;\land\\
&\exists\Myvec s\colon\Myvec\rho.\csem{\Myvec s}\less\Myvec a\;\land
\;\run\infty q{\CN{t\eta}{\Myvec s}}\}
\end{array}$$
\end{DEF}

\begin{REM}\label{rm-csem-closed}
For $t$ a closed term and $\Gamma=\emptyset$ we have $\csemC\Gamma t=\csem t$.
\end{REM}

\begin{prop}\label{prop-csem-subst}
If $s$ has type $\Myvec\sigma\to\iota$ in some context compatible with
$\Gamma$, and $\eta$ is some substitution with $\dom\eta=\dom\Gamma$ such that for all
$x\in\dom\Gamma$ we have $\eta(x)$ closed and
$\csem{\eta(x)}\less\Gamma(x)$, then 
$$\csem{s\eta}\less\csemC\Gamma s$$
\end{prop}
\begin{proof}
Let $\Myvec a\in\sem{\Myvec\sigma}$ and $q\in\csem{s\eta}(\Myvec a)$ be given.
Then there are $\Myvec s\colon\Myvec\sigma$
with $\csem{\Myvec s}\less\Myvec a$ such that $\run\infty q{\CN{s\eta}{\Myvec s}}$.
Together with the assumed properties of $\eta$ this witnesses
$q\in\csemC\Gamma s(\Myvec a)$.
\end{proof}

\begin{lem}\label{lem-csem-app}
If $r$ and $s$ are terms of type $\sigma\to\Myvec\rho\to\iota$ and
$\sigma$, respectively, in some context 
compatible with $\Gamma$, then we have
$$\csemC\Gamma{rs}\less\rep(\csemC\Gamma r\csemC\Gamma s)$$
\end{lem}
\begin{proof}
Let $\Myvec a\in\sem{\Myvec\rho}$ and $q\in\csemC\Gamma{rs}(\Myvec a)$ be
given.
Then
there is $\eta$ with $\forall
x\in\dom\Gamma.\;\csem{\eta(x)}\less\Gamma(x)$ and there are $\Myvec
s\colon\Myvec \rho$ with $\csem{\Myvec s}\less \Myvec a$ and 
$$\run\infty q{\underbrace{\CN{(rs)\eta}{\Myvec
	  s}}_{\rep.\CN{r\eta}{(s\eta,\Myvec s)}}}$$

Hence there is a $q'\in\delta(q,\rep)$ with $\run\infty
{q'}{\CN{r\eta}{(s\eta,\Myvec s)}}$. It suffices to show that for
this $q'$ we have $q'\in\csemC\Gamma r\csemC\Gamma s\Myvec a$.

By Proposition~\ref{prop-csem-subst} we have
$\csem{s\eta}\less\csemC\Gamma s$ and we already have
$\csem{\Myvec s}\less\Myvec a$. So the given $\eta$ together with $s\eta$
and $\Myvec s$ witnesses $q'\in\csemC\Gamma r\csemC\Gamma s\Myvec a$.
\end{proof}

\begin{lem}\label{lem-csem-lam}
Assume that $\lambda x.r$ has type $\sigma\to\Myvec\rho\to\iota$ in some
context compatible with $\Gamma$. Then
$$\csemC\Gamma{\lambda xr}(a) \less\beta(\csemC{\Gamma_x^a}{r})$$
\end{lem}
\begin{proof}
Let $\Myvec a\in\sem{\Myvec\rho}$ and $q\in\csemC\Gamma{\lambda xr}(a,\Myvec
a)$ be given. Then there is an $\eta$ with
$\forall x\in\dom\Gamma$ we have $\eta(x)$ closed and
$\csem{\eta(x)}\less\Gamma(x)$ and there are $s,\Myvec s$ with $\csem s\less
a$ and $\csem{\Myvec s}\less\Myvec a$ such that
$$\run\infty q{\underbrace{\CN{(\lambda xr)\eta}{(s,\Myvec s)}}_{\beta. \CN{r_x[s]\eta}
{\Myvec s}}}$$

So there is a $\tilde q\in\delta(q,\beta)$ with $\run\infty{\tilde q}{\CN{r_x[s]\eta}
{\Myvec s}}$. It suffices to show that $\tilde q\in\csemC{\Gamma_x^a}
r(\Myvec a)$.

By the properties of $\eta$ and since $\csem
s\less a$ we know that for all $y\in\dom{\Gamma_x^a}$ we have
$\csem{\eta(y)}\less\Gamma_x^a(y)$.
This witnesses $\tilde q\in\csemC{\Gamma_x^a} r(\Myvec a)$.
\end{proof}

\begin{lem}\label{csem-var}
$\csemC\Gamma x\less\Gamma(x)$
\end{lem}
\begin{proof}
Assume $x$ of type $\Myvec\rho\to\iota$, let $\Myvec a\in\sem{\Myvec\rho}$
and $q\in\csemC\Gamma x(\Myvec a)$ be given. We have to show
$\Gamma(x)(\Myvec a)$.

Since $q\in\csemC\Gamma x(\Myvec a)$, there is $\eta$ with $\eta(x)\less
a$ and $\Myvec s\colon\Myvec\rho$ with $\csemC\Gamma{\Myvec s}\less\Myvec a$
and $\run\infty q{\CN{\underbrace{x\eta}_{\eta(x)}}{\Myvec s}}$.

But then $\Myvec s$ witness that $q\in\csem{\eta(x)}(\Myvec
a)\subset\Gamma(x)(\Myvec a)$ where the last subset relation holds since
$\csem{\eta(x)}\less\Gamma(x)$. 
\end{proof}


\begin{Th}\label{th-sound}
$\fproof n\Gamma{\csemC\Gamma t}t{\rho}$
\end{Th}
\proof
Induction on $n$, cases on $t$. Trivial for $n=0$. So let $n>0$. We
distinguish cases according to $t$
\begin{enumerate}[$\bullet$]
\item Case $rs^\sigma$. By induction hypothesis
  $\fproof{n-1}\Gamma{\csemC\Gamma r} r {\sigma\to\rho}$ and
  $\fproof{n-1}\Gamma{\csemC\Gamma s} s \sigma$. Moreover, by
  Lemma~\ref{lem-csem-app} $\csemC\Gamma{rs}\less\rep(\csemC\Gamma
  r\csemC\Gamma s)$. Hence $\fproof
  n\Gamma{\csemC\Gamma{rs}}{rs}\rho$.
\item Case $\lambda x^\sigma r$. By induction hypothesis we have for
  all $a\in\sem\sigma$ that
  $\fproof{n-1}{\Gamma_x^a}{\csemC{\Gamma_x^a}r}r\rho$. By
  Lemma~\ref{lem-csem-lam} we have
 $\csemC\Gamma{\lambda xr}(a)\less\beta(\csemC{\Gamma_x^a}r)$.  

  Hence $\fproof n\Gamma{\csemC\Gamma{\lambda xr}}{\lambda
  xr}{\sigma\to\rho}$.
\item Case $x$. By Lemma~\ref{csem-var} we have $\csemC\Gamma
  x\less\Gamma(x)$ and hence
$\fproof n\Gamma{\csemC\Gamma x}x\rho$.
\item Case $t=\term$ a terminal symbol.
We have to show
$\fproof n\Gamma{\csemC\Gamma\term}\term{\iota\to\iota}$.

So, let $\Myvec S\in\sem{\Myvec\iota}$ and
$q\in\csemC\Gamma\term(S)$. Hence there are $\Myvec s$ of type
$\iota$ with $\csem {s_i}\less S_i$ and
$\run\infty q{\underbrace{\CN\term{\Myvec s}}_{\term(\Myvec{s^\beta})}}$.

So there is ${(\tilde q_1,\ldots,\tilde
  q_{\ar\term},\ast,\ldots,\ast)}\in\delta(q,\term)$ 
with $\run\infty
{\tilde q_i}{s_i^\beta}$. But then $\tilde q_i\in\csem {s_i}\subset S_i$.\qed
\end{enumerate}

\begin{cor}\label{cor-compl}
If $t\colon\iota$ is closed and of ground type then 
$\fproof n\emptyset{\{q\mid\run\infty q{t^\beta}\}}t\iota$.
\end{cor}
\begin{proof}
By Remarks~\ref{rm-csem-closed} and~\ref{rm-csem-base} we have
$\csemC\emptyset t=\csem t = \{ q\mid\run\infty q{t^\beta}\}$. So the
claim follows from Theorem~\ref{th-sound}.
\end{proof}

Finally, let us sum up what we have achieved.
\begin{cor}\label{cor-main-result}
For $t$ a closed regular lambda term, and $q_0\in Q$ it is decidable
whether $\run\infty{q_0}{t^\beta}$.
\end{cor}
\begin{proof}
By Proposition~\ref{prop-proof-dec} it suffices to show that
$\fproof\infty\emptyset{\{q_0\}}t\iota$ holds, if and only if
$\run\infty{q_0}{t^\beta}$.

The ``if''-direction follows from
Corollary~\ref{cor-compl} and the weakening provided by
Remark~\ref{rem-proof-weak}. The ``only if''-direction is provided by
Corollary~\ref{cor-sound}.
\end{proof}

Note that, since there are only finitely many ways to extend a proof
of level $n$ to a proof of level $n+1$ and all proofs of level $n+1$
come from a proof of level $n$ the corollary implies, by K\"onig's
Lemma, that $\run\infty q{t^\beta}$ implies
$\fproof\infty\emptyset{\{q\}}t\iota$.

%
%
%
%
%

\section{Model Checking}\label{sEc:model-checking}

\begin{Th}
Given a tree $\mathcal T$ defined by an arbitrary recursion scheme (of
arbitrary level) and a property $\varphi$ expressible by a trivial
automaton, it is decidable whether $\mathcal T\models\varphi$.
\end{Th}
\begin{proof}
Let $t$ be the infinite lambda-tree associated with the recursion
scheme. Then $t$ is effectively given as a regular closed lambda term
of ground type and $\mathcal T$ is the normal form of $t$.

Let $\A_\varphi$ be the automaton (with initial state $q_0$)
describing $\varphi$. By keeping the state when reading a $\rep$ or
$\beta$ it can be effectively extended to an automaton $\A$ that works
on the continuous normal form, rather than on the usual one.
So $\mathcal
T\models\varphi\Leftrightarrow\run\infty{q_0}{t^\beta}$. The latter,
however, is decidable by Corollary~\ref{cor-main-result}.
\end{proof}

\begin{REM}
As shown in Section~\ref{sEc:autom-new}, the above theorem is
in particular applicable to $\mathsf{CTL}$-properties built from
letters, conjunction, disjunction, ``next'', and ``globally''.
\end{REM} 

\begin{REM}
As discussed after Proposition~\ref{prop-proof-dec} the complexity is
fixed-parameter non-deterministic linear time in the size of the recursion
scheme, if we consider $\varphi$ and the allowed types as a parameter.
\end{REM}

Finally, looking back at the technical development, it is not clear to
the author, whether this approach can be extended in a smooth way to work 
for arbitrary automata, as opposed to only trivial ones.
It is tempting to conjecture that appropriate annotations of the proofs
with priorities could extend the concept to parity automata (and hence the full
of Monadic Second Order). However, all the ways that seemed obvious to the 
author failed.

One technical problem is that several paths might lead to the same state at the same
node, but with different priorities visited so far. A more fundamental problem is the
way the runs are constructed in the proofs throughout this article; we're given a run
by induction hypothesis and add a move at its \emph{beginning}. As all acceptance conditions
ignore finite prefixes, all the promises to visit some state eventually are pushed in the 
future indefinitely. So, some promise on how long it will take for some promised event to happen
seems to be needed in the annotations, at least if we want these global conditions to fit with
our local arguments. It is not clear to the author whether and how this can be achieved.

}\vskip-20 pt
\end{document}